\documentclass[twocolumn,final,aps,prb,showpacs,superscriptaddress,amsmath,amssymb,amsfonts,floatfix,longbibliography]{revtex4-1}
\usepackage{epsfig}
\usepackage[dvipsnames,usenames]{color}
\usepackage{color}
\usepackage{amsmath}
\usepackage{comment}
\usepackage{array}
\usepackage{multirow}
\usepackage{tabularx}
\usepackage{float}
\usepackage{ulem}
\usepackage[colorlinks=true,bookmarks=true, pdfborder={0 0 0},linkcolor=blue,urlcolor=blue,    breaklinks=true,bookmarksnumbered=true]{hyperref}

\newcommand{\py}[1]{{\color{black}#1}}

\emergencystretch=\maxdimen
\begin{document}
\title{Interplay between Hubbard interaction and charge transfer energy in three-orbital Emery model: implication on cuprates and nickelates}
\author{Yan Peng}
\affiliation{School of Physical Science and Technology, Soochow University, Suzhou 215006, China}
\author{Mi Jiang}
\email[]{jiangmi@suda.edu.cn}
\affiliation{School of Physical Science and Technology, Soochow University, Suzhou 215006, China}
\affiliation{Jiangsu Key Laboratory of Frontier Material Physics and Devices, Soochow University, Suzhou 215006, China}

\begin{abstract}
We use the numerically unbiased determinant quantum Monte Carlo (DQMC) method to systematically investigate the three-orbital Emery model in the normal state in a wide range of local interactions, charge transfer energy, and doping levels. We focus on the influence of the onsite Hubbard $U_{dd}$ and \py{the} charge transfer energy scale $\epsilon_p$ on the electronic properties via the orbital occupancies, local moments, spin correlations, and spectral properties. 
Rich features of the orbital-resolved local and momentum-dependent spectra are revealed to associate with the possible Zhang-Rice singlet (ZRS) breakdown reflected by the peak splitting near the Fermi level in the heavily overdoped regime. 
Moreover, the pseudogap features at \py{a} small charge transfer energy scale (relevant to cuprates) are shown to diminish at larger $\epsilon_p$, which implies the weakening or absence of the pseudogap in the infinite-layer nickelates.
Besides, an optimal value of $\epsilon_p$ is identified for maximizing the antiferromagnetic (AFM) spin correlations. Our large-scale simulations provide new insights on the well-established Emery model, particularly in the regime of heavily overdoped and/or large charge transfer energy scale. 
\end{abstract}

\maketitle

\section{Introduction}
Cuprate high-temperature superconductors (SC) have remained a subject of extensive research since their discovery in the 1980s~\cite{bednorz_possible_1986}. Other emergent phenomena such as pseudogap, stripe phase, and strange metal behavior~\cite{wu_pseudogap_2018, tajima_correlation_2024, ponsioen_superconducting_2023, mai_fluctuating_2024} render  \py{their} underlying physical mechanism even more elusive. Due to the quasi-two-dimensional structure of Cu-O planes and strong local interaction on Cu sites, an effective single-orbital model which originates from a more involved three-orbital Emery model has been proposed to explain the low-energy physics dominated by the well-known Zhang-Rice singlet (ZRS)~\cite{zhang_effective_1988}. However, the omission of oxygen degrees of freedom makes the regime of its validity unclear, especially based on the experimental fact that the cuprates are charge transfer insulators (CTI) rather than Mott Hubbard insulators (MHI) in essence ~\cite{zaanen_band_1985}. Recent experimental and theoretical studies on cuprates have also challenged the applicability of the single-orbital model in the overdoped regime~\cite{kim_optical_2021, chen_doping_2013, li2022strongly}. Hence, the three-orbital Emery model~\cite{emery_theory_1987} explicitly including $3d_{x^2-y^2}$, $2p_x$, and $2p_y$ orbitals in Cu-O plane, is  \py{closer} to the realistic physical picture without assuming the existence of ZRS in advance.

\begin{figure}[t!]
\psfig{figure=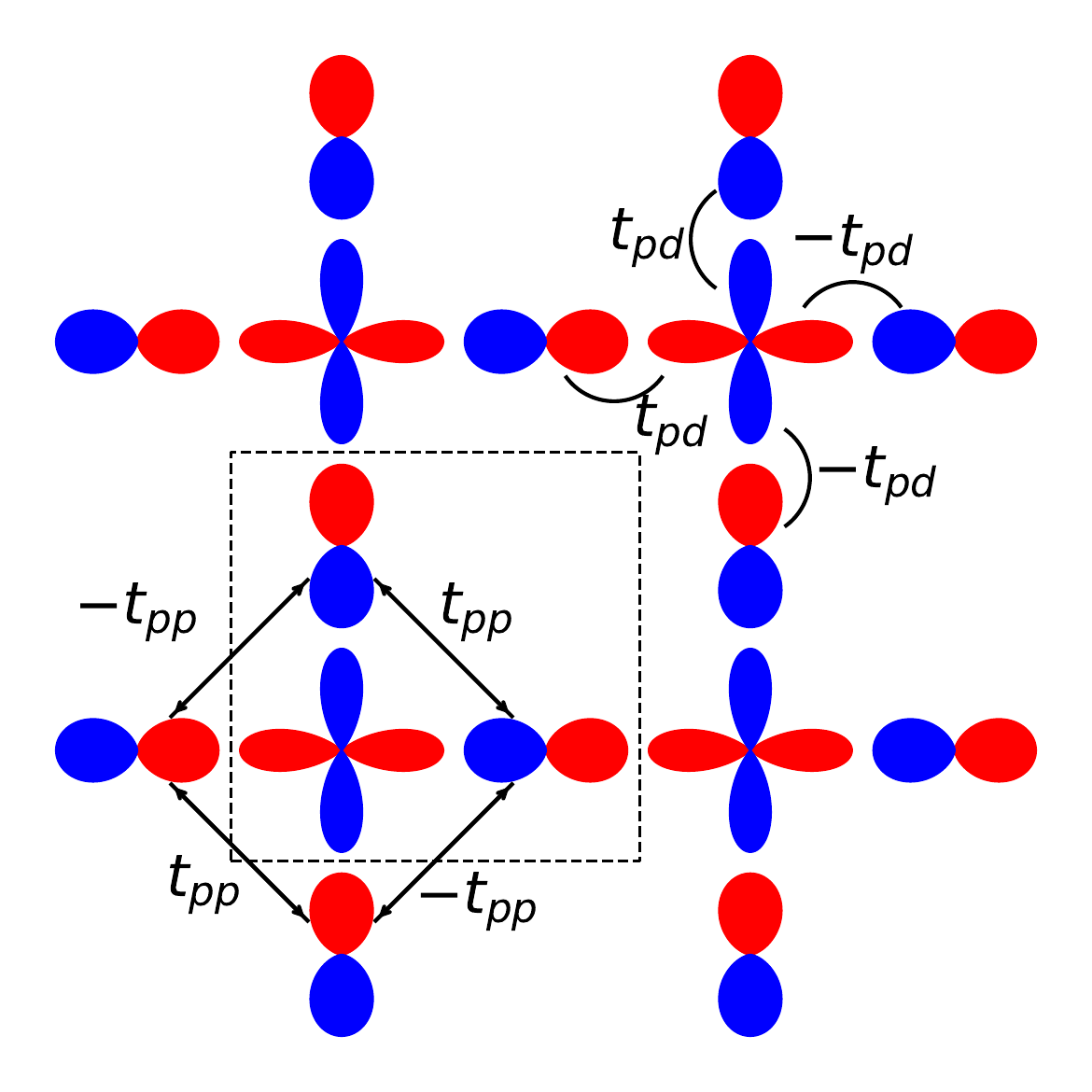,width=0.40\textwidth,clip}
\caption{A schematic illustration of a Cu-$d_{x^2-y^2}$ orbital and its four nearest-neighbor O-$p_{x/y}$ orbitals. Red (blue) color indicates positive (negative) phase factor. The unit cell is outlined by the dashed box. \py{The phase convention of the hopping parameters is defined in the hole language.}}
\label{3bandmodel}
\end{figure}

The three-orbital model has been applied to investigate cuprate SC since its discovery and \py{has} already been shown to be \py{a} plausible framework. For example, previous work~\cite{scalettar_antiferromagnetic_1991,maier_magnetic_1999} uncovered the strong suppression of the antiferromagnetic state upon doping by both exact diagonalization (ED) and dynamical mean-field theory (DMFT). Contemporaneously, Guerrero \textit{et al.}~\cite{guerrero_quantum_1998} used the constrained-path Monte Carlo method to demonstrate that the $d$-wave pairing correlations dominate the extended $s$-wave. Medici \textit{et al.}~\cite{de_medici_correlation_2009} studied the doping asymmetry of ZRS using DMFT. They pointed out that the cuprates are in an intermediate correlation regime and  cast doubt on the validity of the ZRS approximation when the charge transfer energy enters into the large regime. \py{In recent years, owing to improved computational capabilities,} there is growing support for the emergence of superconductivity~\cite{ponsioen_superconducting_2023}, pseudogap~\cite{tajima_correlation_2024, wu_pseudogap_2018}, and density wave orders~\cite{xu_coexistence_2024, mai_fluctuating_2024, jiang_pair_2023} within the three-orbital model. Another key motivation arises from the recently discovered infinite-layer nickelate high-${\mathrm{T_c}}$ superconductors~\cite{li2019superconductivity}, which have a larger charge transfer energy~\cite{mi20prl,kapeghian2020electronic, kreisel2022superconducting} than cuprates and are closer to a competing regime of Mott Hubbard versus charge transfer dominance~\cite{ding2024cuprate, lechermann2024oxygen}, in spite of the claimed importance of other orbitals like interstitial $s$ orbital~\cite{gu2020substantial, chen2022dynamical, li2025observationelectridelikesstates}.

In this work, we adopt \py{the DQMC} method to systematically investigate the Emery model in the normal state in a wide range of local interactions, charge transfer energy, and doping levels \py{on} a lattice size larger than before~\cite{georges_dynamical_1996, de_medici_correlation_2009, weisse2008exact, lin1988pairing}. In particular, \py{in addition to} a broad range of the Cu local interaction $U_{dd}$, we extend the O site energy $\epsilon_p$ from the conventional charge transfer regime relevant to cuprates to the Mott Hubbard regime in the Zaanen-Sawatzky-Allen scenario~\cite{zaanen_band_1985}. Besides, we push the doping level to highly overdoped regime. 

This paper is organized as follows: 
Section~\ref{Model and Method} presents the three-orbital Emery model and the basic principle of DQMC methodology. Then in Section~\ref{sec:results}, we first analyze the density distribution and spectral functions to investigate the quasiparticle behavior and band renormalization. Subsequently, we examine the magnetic correlations to capture the essential collective behavior under different $U_{dd}$ and $\epsilon_p$ combinations.
Finally, Section~\ref{sec:conclusion} summarizes our work.

\section{Model and Method}
\label{Model and Method}
The three-orbital Emery model involving the Cu-$3d_{x^2 - y^2}$, O-$2p_x$, and O-$2p_y$ orbitals, with all onsite interactions taken into account, reads as
\begin{align}
\hat{H}&=\hat{E^s}+\hat{K}^{pd}+\hat{K}^{pp}+\hat{U} \nonumber \\ 
    \hat{E^s}&=(\epsilon_d-\mu)\sum_{i\sigma}\hat{n}^d_{i\sigma}+(\epsilon_p-\mu)\sum_{j\sigma}\hat{n}^p_{j\sigma} \nonumber \\ 
    \hat{K}^{pd}&=\sum_{\langle ij \rangle\sigma}t^{ij}_{pd}(\hat{d}^\dagger_{i\sigma}\hat{p}_{j\sigma}+h.c.) \nonumber \\
    \hat{K}^{pp}&=\sum_{\langle jj^{\prime} \rangle\sigma}t^{jj^{\prime}}_{pp}(\hat{p}^\dagger_{j\sigma}\hat{p}_{j^{'}\sigma}+h.c.) \nonumber \\
    \hat{U}&=U_{dd}\sum_{i}\hat{n}_{i\uparrow}\hat{n}_{i\downarrow}+U_{pp}\sum_{j}\hat{n}_{j\uparrow}\hat{n}_{j\downarrow} ,
\end{align}
where $\hat{E}^s$ represents the onsite energies of $3d_{x^2-y^2}$ orbital at site $i$ and $2p_{x/y}$ orbital of site $j$, where $\hat{n}^d_{i\sigma}$ ($\hat{n}^p_{j\sigma}$) is the hole density operator for the $d$ ($p$) orbital, whose on-site energy is $\epsilon_d$ ($\epsilon_p$). The chemical potential $\mu$ controls the total occupancy. The kinetic energy terms $\hat{K}^{pd}$ and $\hat{K}^{pp}$ describe the nearest-neighbor (NN) Cu-O and O-O hoppings, denoted by $\langle ij \rangle$ or $\langle jj^{\prime} \rangle$, in corresponding order. Specifically, we choose the hole language so that $\hat{d}^{\dagger}_{i\sigma}$ ($\hat{d}_{i\sigma}$) creates (annihilates) a hole with spin $\sigma$ on a $d$ orbital at site $i$. The same applies to the $p$ orbital operators. Besides, the hopping integrals $t_{pd}^{ij}$ and $t_{pp}^{jj^{\prime}}$ take the convention as
\begin{align}
    t_{pd}^{ij}&=t_{pd}(-1)^{\eta_{ij}} \nonumber \\
    t_{pp}^{jj^{\prime}}&=t_{pp}(-1)^{\xi_{jj^{\prime}}}
\end{align}
with the phase convention $\eta_{ij} = 1$ for $j = i + \frac{\hat{x}}{2}$ or $j = i - \frac{\hat{y}}{2}$ and $\eta_{ij} = 0$ for $j = i - \frac{\hat{x}}{2}$ or $j = i + \frac{\hat{y}}{2}$, where the vectors $\hat{x}$ and $\hat{y}$ are the in-plane unit cell basis vectors. Similarly, $\xi_{jj^{\prime}} = 1$ for $j^\prime = j + \frac{\hat{x}}{2} + \frac{\hat{y}}{2}$ or $j^\prime = j - \frac{\hat{x}}{2} - \frac{\hat{y}}{2}$ and $\xi_{jj^{\prime}} = 0$ for $j^\prime = j + \frac{\hat{x}}{2} - \frac{\hat{y}}{2}$ or $j^\prime = j + \frac{\hat{x}}{2} - \frac{\hat{y}}{2}$. Both the unit cell and the sign of the hopping amplitude are illustrated in Figure~\ref{3bandmodel}. Due to gauge invariance, this choice of signs is not unique as mentioned in other studies~\cite{fischer_mean-field_2011, mao_non-fermi-liquid_2024, mai_pairing_2021}.


The interaction term $\hat{U}$ describes the onsite repulsion on the holes of $d_{x^2-y^2}$ or $p_{x/y}$ orbitals. In the limit of $t_{pp} = 0$, the amplitude of $U_{dd}$ together with the charge transfer energy $\Delta=\epsilon_p-\epsilon_d$ significantly affect\py{s} the insulating behavior of the ground state at half-filling~\cite{sawatzky_explicit_2016, andersen1995lda}. For $U_{dd} < \Delta$ it is a Mott Hubbard insulator\py{,} whereas for $U_{dd} > \Delta$ it is a charge transfer insulator~\cite{zaanen_band_1985}. Cuprates fall into the latter category. 

According to the canonical parameter set~\cite{mai_fluctuating_2024, kung_characterizing_2016}, we rescale our parameter set as $t_{pd}=1.0$, $t_{pp}=0.4$, $\epsilon_d=0$ and $U_{dd}$ from 4.0 to 8.0, $\epsilon_p$ from 2.0 to 6.0 for convenience\py{. Therefore, }$t_{pd}=1.0$ serves as the energy unit. In the hole language, half-filling is defined as $\langle n_{\mathrm{tot}} \rangle = 1$ and hole (electron) doping corresponds to $\langle n_{\mathrm{tot}} \rangle > 1$ ($<1$).

We remark that the omission of $U_{pp}$ alleviates the sign problem, allowing us to access larger lattice sizes up to $8\times8$ (and even $12\times12$ not shown here) than previous works~\cite{kung_characterizing_2016, mai_pairing_2021} at reasonably low temperatures. Unless otherwise specified, the results shown below are for $8 \times 8$ lattice and the inverse temperature $\beta t_{pd}=10.0$. Apart from the investigation on the interplay between $U_{dd}$ and $\epsilon_p$, we also examined the role of $U_{pp}$ in a smaller lattice size, which was found to \py{have little effect on} the spin-spin correlations~\cite{kung_characterizing_2016}. Nonetheless, we will show that the low\py{-}energy excitations obtained from the spectral functions show strong dependence on $U_{pp}$ in the hole doping regime.


To fully take all the energy scales into account  on \py{an} equal footing, we use the well\py{-}established numerical technique of finite temperature \py{DQMC}~\cite{blankenbecler_monte_1981}. As a celebrated computational method, DQMC provides a numerically unbiased solution in the presence of strong correlations.

In order to deal with the \textit{ill-posed} problem caused by the inversion of the fermionic imaginary-time Green\py{'s} function to obtain the spectral function via
\begin{align}
    G(\mathbf{k}, \tau)=\int_{-\infty}^{+\infty}\frac{d\omega}{2\pi}\frac{e^{-\omega\tau}A(\mathbf{k}, \omega)}{1+e^{-\beta\omega}}
    \label{MaxEnt}
\end{align}
we adopt the maximum entropy analytic continuation method (MaxEnt) to extract the least
biased spectral function from all the feasible solutions~\cite{wu_maximum_2012}. We examine both the density of states (DOS) and the orbital- and momentum-resolved spectral function $A_\alpha(\mathbf{k}, \omega)$. Based on the equal-time or the time-dependent Green\py{'s} function, substantial physical quantities, such as the spin-spin correlation and the single-particle spectral functions, allow us to access a thorough and comprehensive understanding of the Emery model.

\section{Results}
\label{sec:results}

\subsection{Orbital-resolved density distribution}

\begin{figure}
\psfig{figure=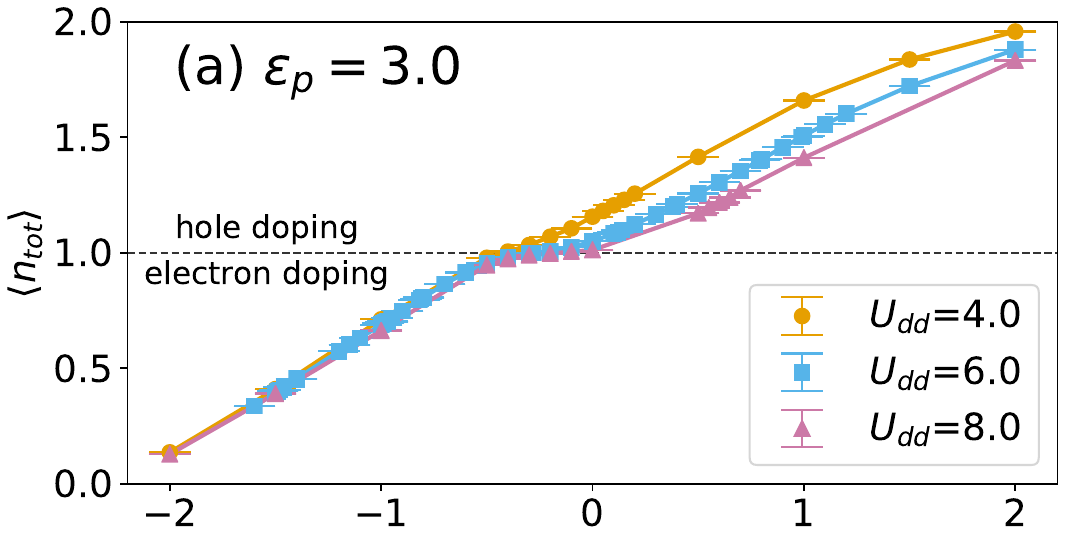,width=.45\textwidth, clip} 
\psfig{figure=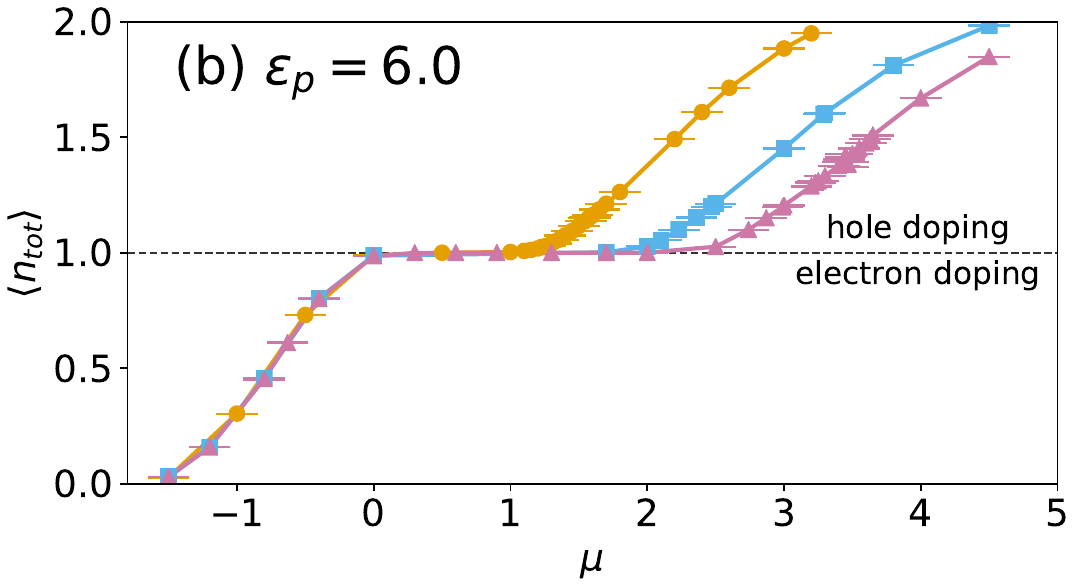,width=.45\textwidth, clip} 
\caption{Total filling $\langle n_{\mathrm{tot}} \rangle$ versus the chemical potential $\mu$ for fixed $\epsilon_p=$ 3.0 (a) \py{and 6.0 (b)} with varying $U_{dd}$. \py{The dashed line represents half-filling, with the hole-doped and electron-doped regions labeled above and below the line respectively.}}
\label{ntt_mu}
\end{figure}

We start by the variation of the total density $\langle n_{\mathrm{tot}} \rangle$ with $\mu$, $U_{dd}$, and $\epsilon_p$. As illustrated in Fig.~\ref{ntt_mu}(a), \py{for a cuprate-like value of $\epsilon_p = 3.0$, due to the dynamical hopping through the Oxygen sublattice and the finite temperature, the system is insufficient to open an energy gap, indicating a metallic behavior at half-filling when $U_{dd}=4.0$. A larger $U_{dd}=6.0, 8.0$ is required to opening an energy gap. In contrast, a larger $\epsilon_p = 6.0$ induces clear gaps shown in Fig.~\ref{ntt_mu}(b), implying the enlarged effective interaction with increasing $\epsilon_p$.}

\begin{figure}
\psfig{figure=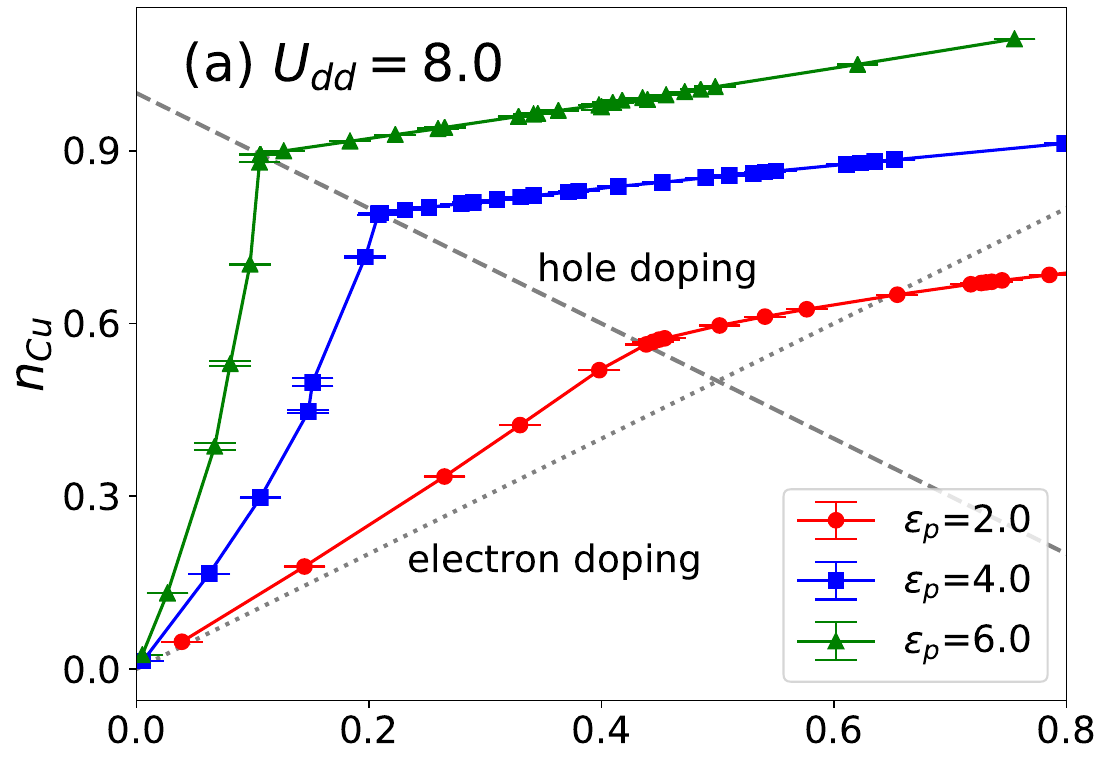,height=5.2cm,width=.45\textwidth, clip} 
\psfig{figure=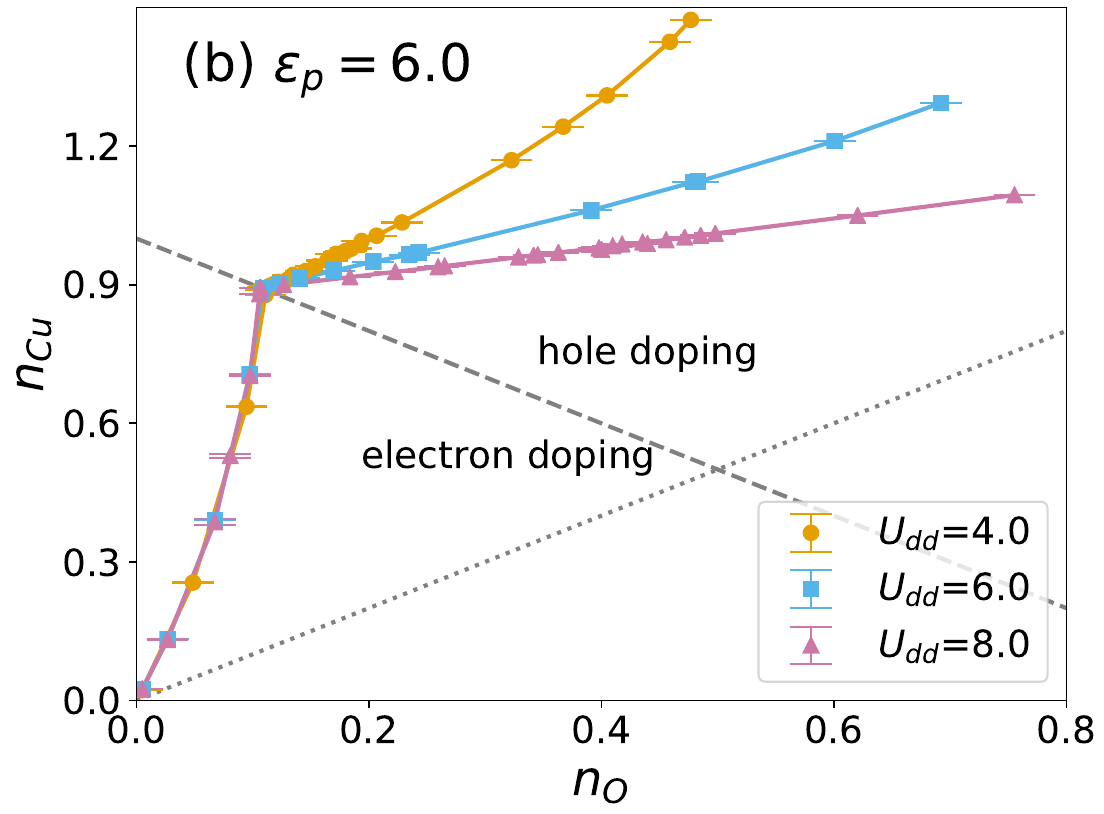,height=5.2cm,width=.45\textwidth, clip} 
\caption{The hole density on Cu orbital $\langle n_{\mathrm{Cu}} \rangle$ versus that on O orbital $\langle n_{\mathrm{O}} \rangle$ with $\epsilon_p$ or $U_{dd}$ being fixed. The dashed line denotes the half-filling $\langle n_{\text{tot}} \rangle = 1.0$ case. \py{The hole-doped and electron-doped regions are labeled at the upper-right and lower-left sides of the line, respectively.} The dotted line indicates $\langle n_{\mathrm{Cu}} \rangle = \langle n_{\mathrm{O}} \rangle$, i.e. equal occupancies on Cu and O.}
\label{nCu_nO}
\end{figure}

In order to further reveal the electron-hole asymmetry~\cite{armitage2010progress}, Fig.~\ref{nCu_nO} shows the mutual dependence between $\langle n_{\mathrm{Cu}} \rangle$ and $\langle n_\mathrm{O} \rangle $ with varying $U_{dd}$ or $\epsilon_p$. The gray dashed line denotes the half-filling case. We emphasize that here $\langle n_\mathrm{O} \rangle $ is the sum of the hole densities of $\mathrm{O_x}$ and $\mathrm{O_y}$ orbitals. The same applies to the O spectra discussed in the following section. \py{Panel} (a) indicates that the turning point at half-filling becomes sharper and the slope on the electron doping side grows quickly, which signifies preferential doping onto Cu. Obviously, as a direct result of larger $\epsilon_p$, the critical $\langle n_\mathrm{O} \rangle $ of the turning point shifts to smaller values and the Cu–O density asymmetry becomes more pronounced.
In panel (b), at a fixed large $\epsilon_p=6.0$, similar to Fig.~\ref{ntt_mu}, $U_{dd}$ has almost no effect on the charge distribution on the electron-doping side and pushes more holes into O orbitals. These behaviors vividly demonstrate the influence of $U_{dd}$ or $\epsilon_p$ on the charge distribution in the three-orbital model. Specifically, $U_{dd}$ prohibits double occupancy on the Cu orbitals, while a lower $\epsilon_p$ than $U_{dd}$ allows holes to avoid energy penalties by occupying the O orbitals, hence leading to a strong asymmetry between electron and hole doping.

\subsection{Orbital-resolved local density of states (LDOS)}

\begin{figure*}
\psfig{figure=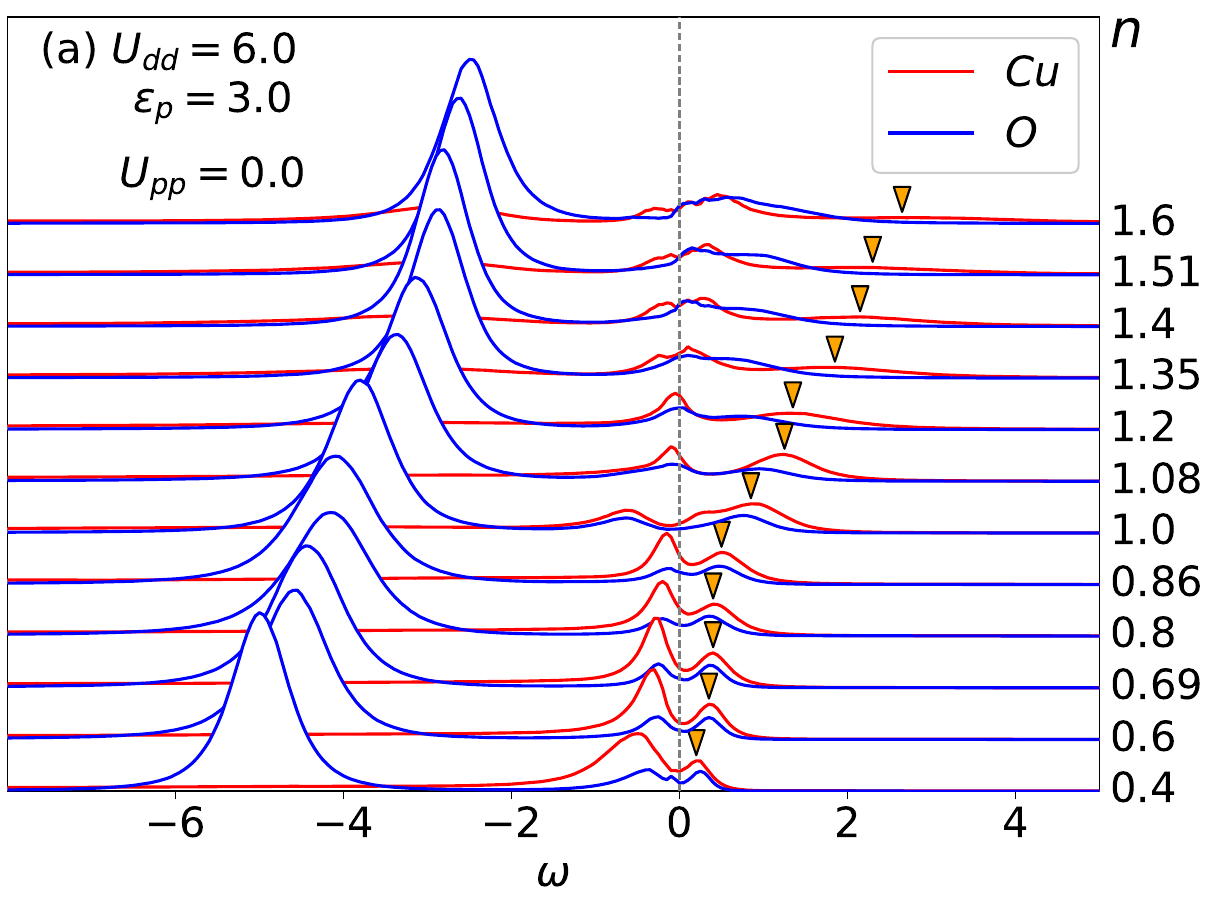,width=.325\textwidth,clip}
\psfig{figure=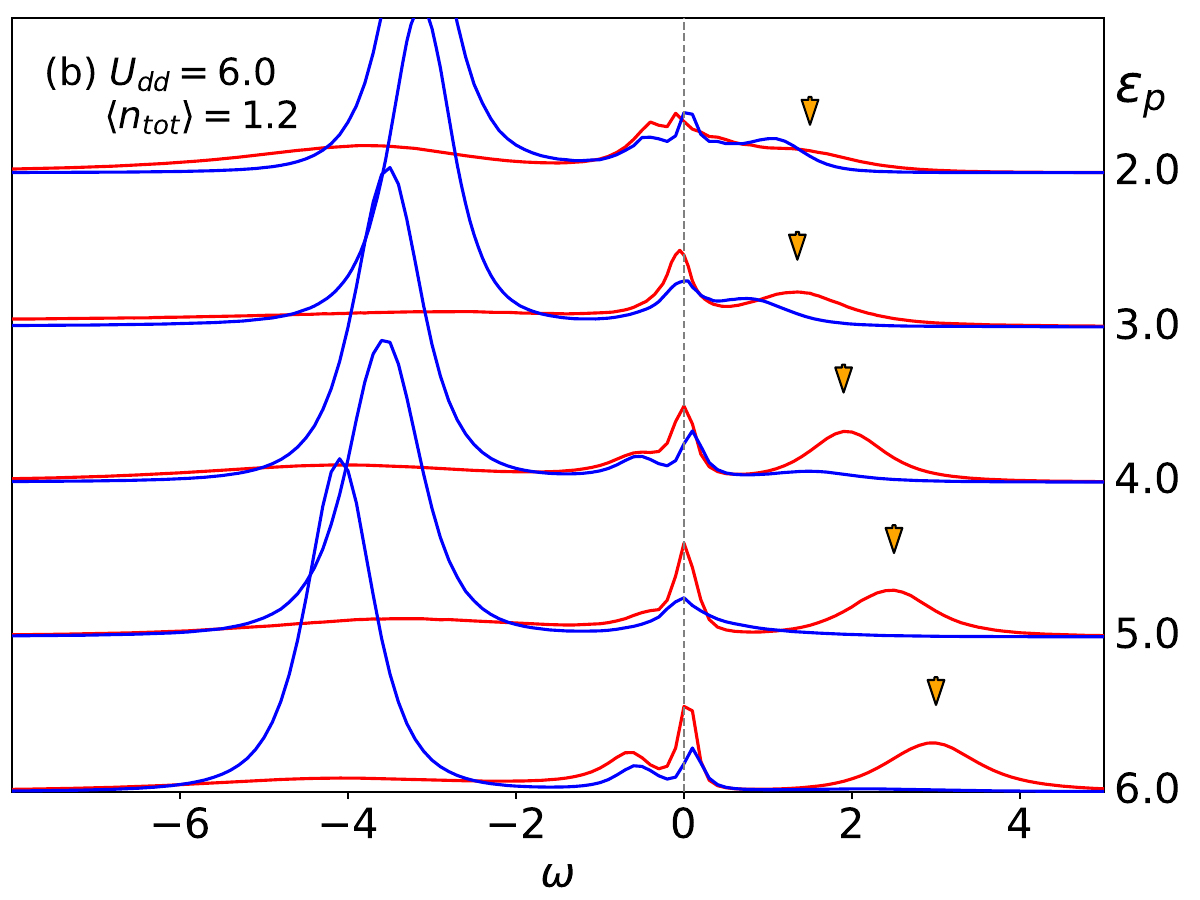,width=.325\textwidth,clip}
\psfig{figure=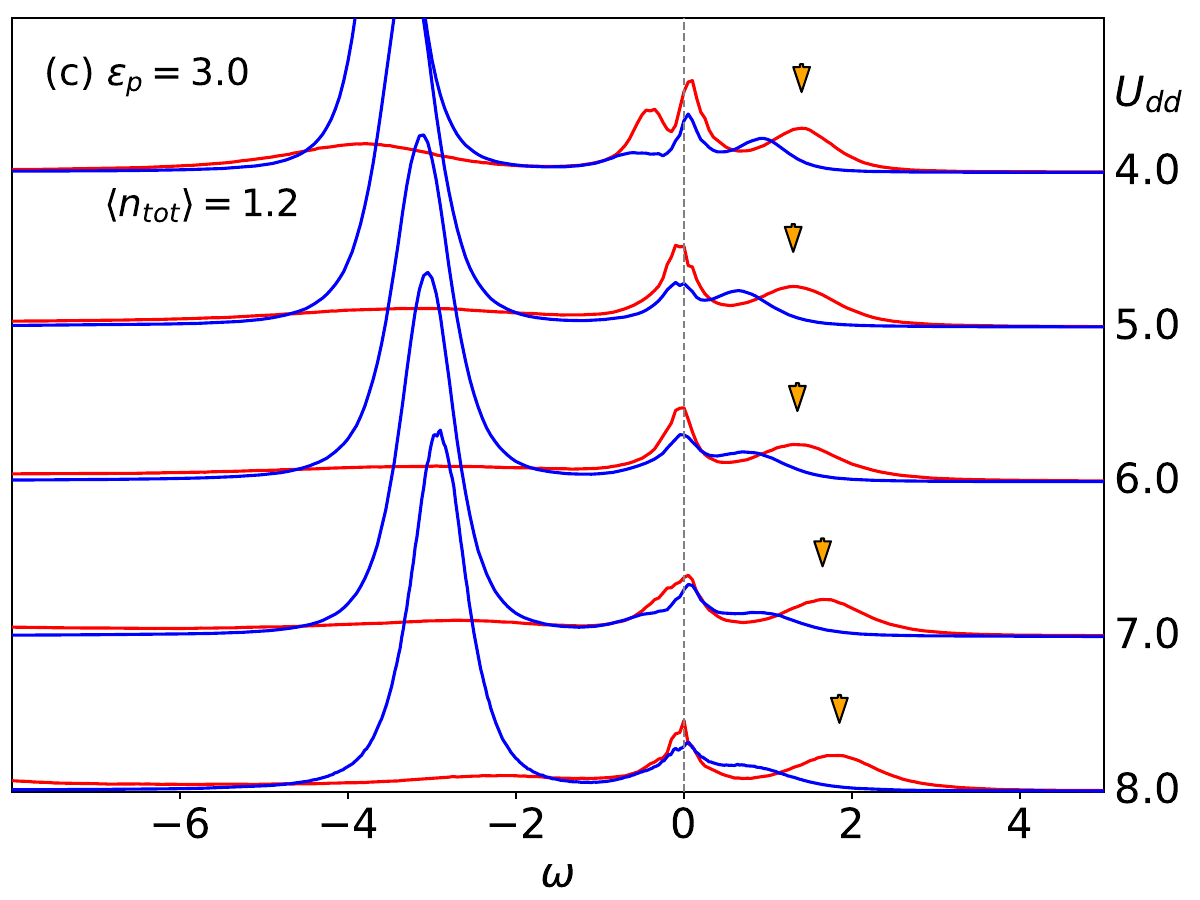,width=.325\textwidth,clip}


\psfig{figure=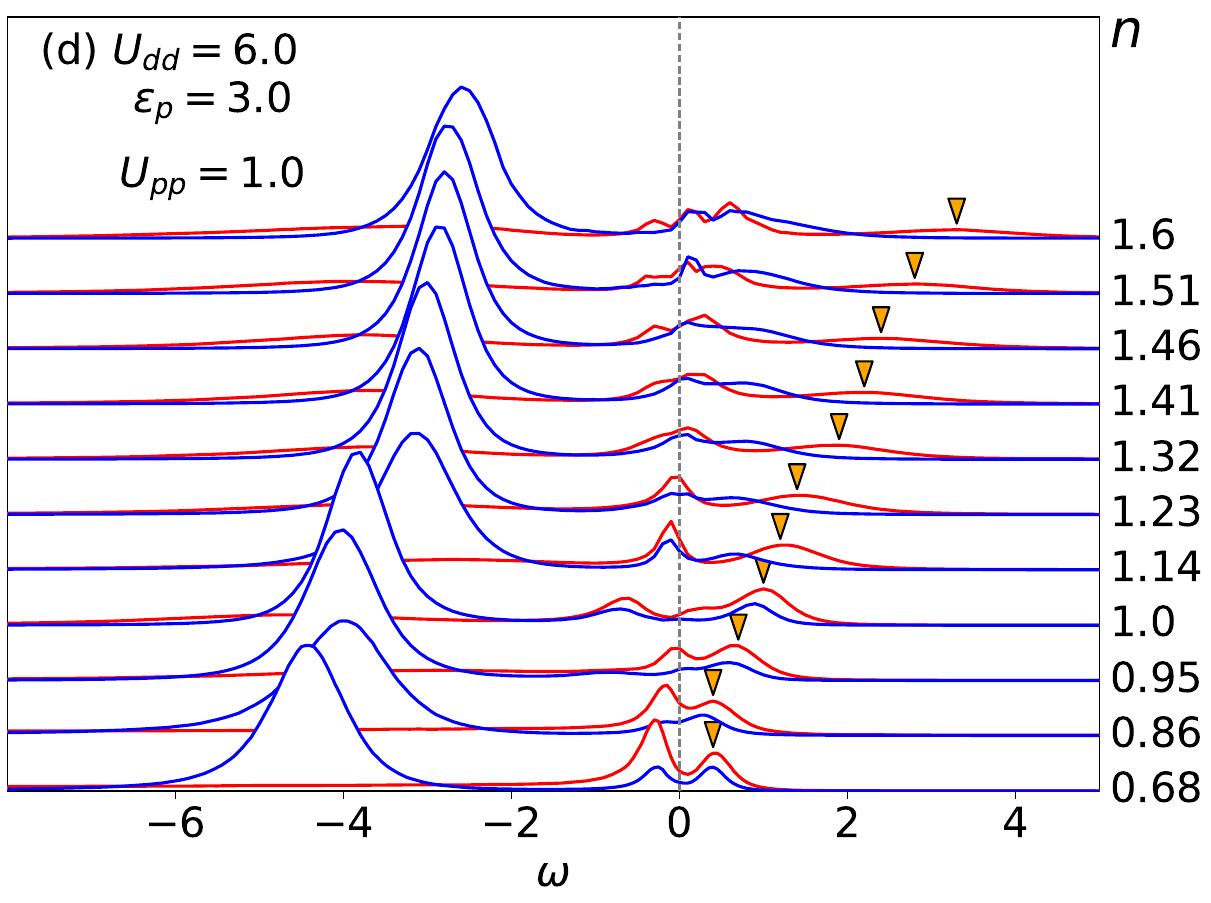,width=.325\textwidth,clip}
\psfig{figure=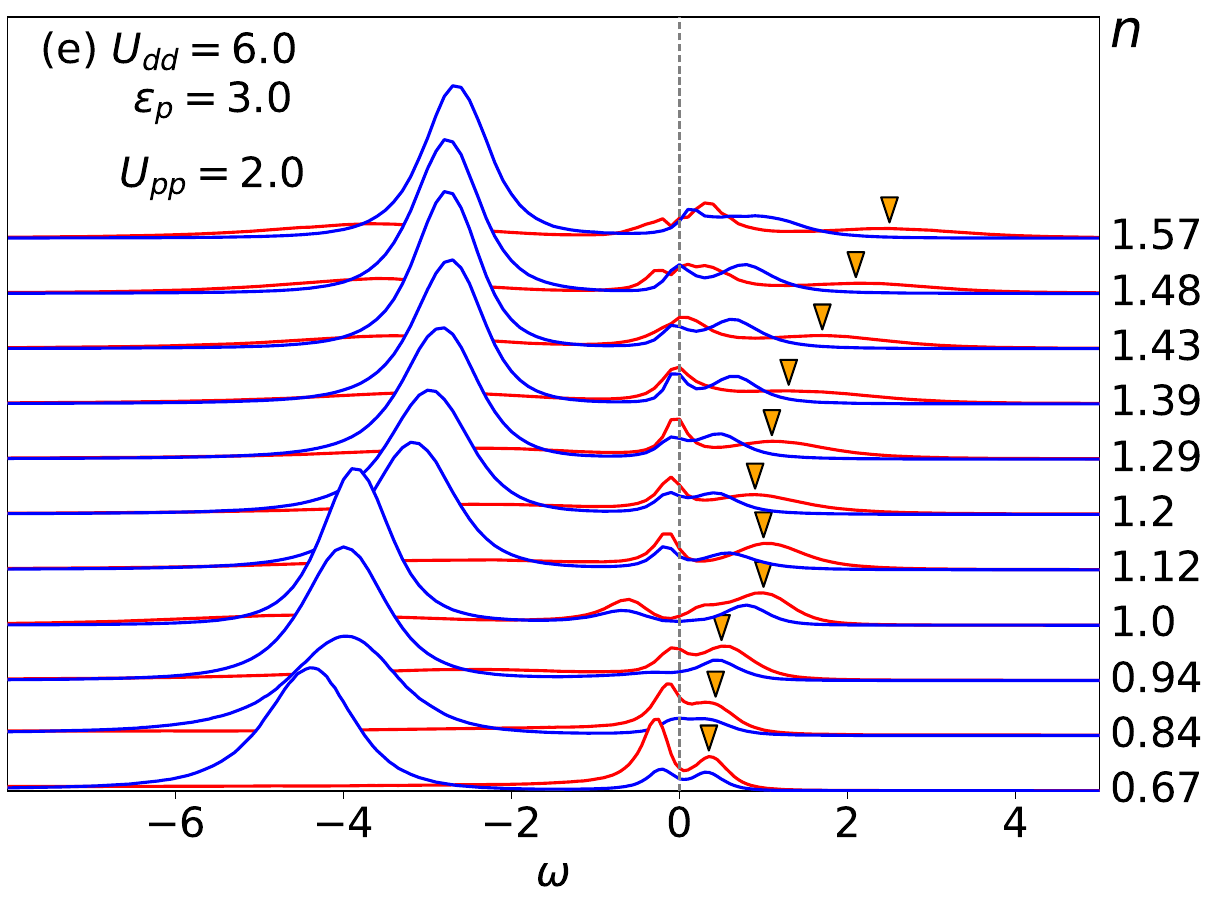,width=.325\textwidth,clip}
\psfig{figure=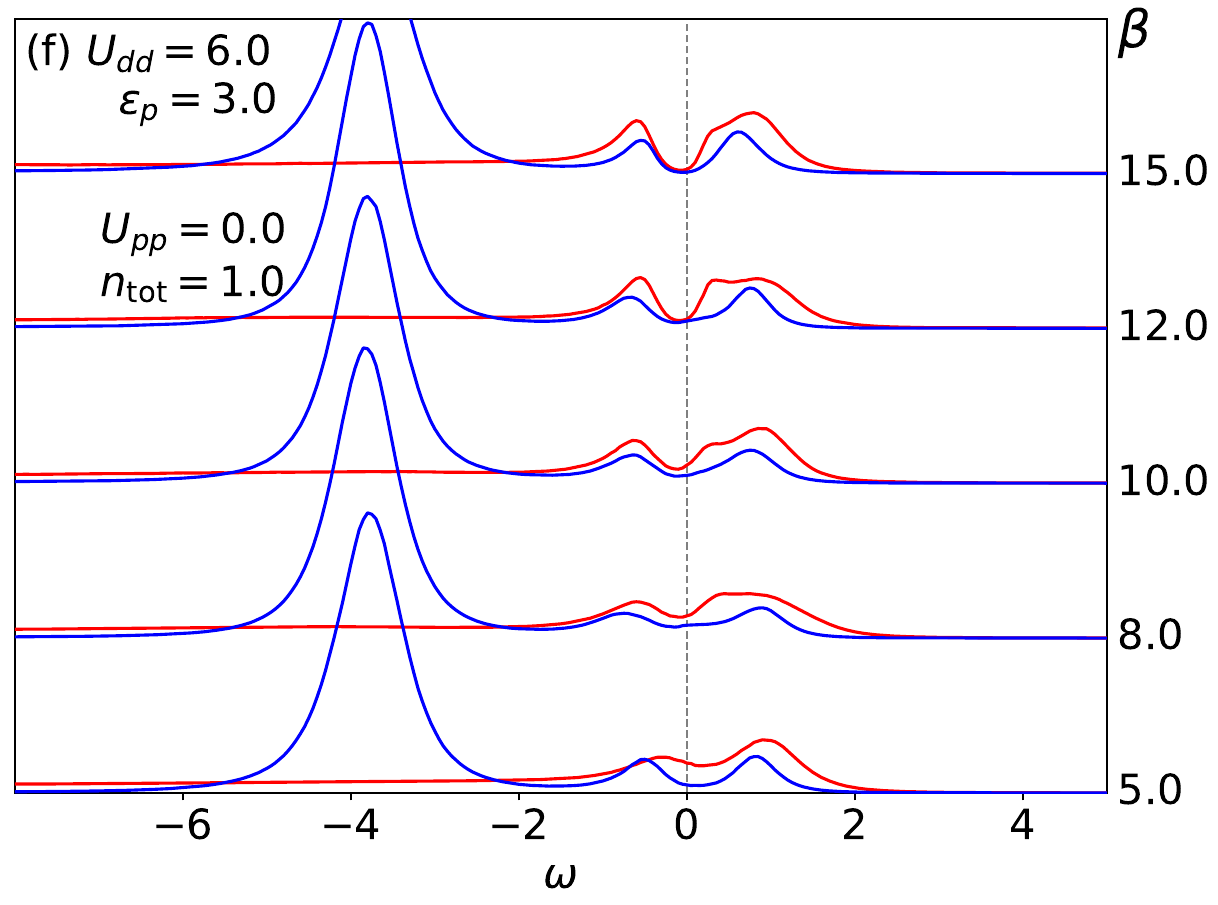,width=.325\textwidth,clip}
\caption{Local density of states (LDOS) as a function of \py{(a, d, e) doping level, (b) $\epsilon_p$, (c) $U_{dd}$, and (f) $\beta$}. The varied parameter values are indicated to the right of each panel. The spectra of O orbital is summed over the $x$- and $y$-direction. The orange triangles denote the location of the UHB.}

\label{LDOS}
\end{figure*}

The orbital-resolved spectral functions can be obtained via analytic continuation of the imaginary-time Green\py{'s} functions, which encode rich information about quasiparticle excitations and gap features in corresponding orbitals.
Fig.~\ref{LDOS}(a)–(c) display the impact on LDOS caused by doping, $\epsilon_p$ and $U_{dd}$, separately. In Fig.~\ref{LDOS}(a), we display a fairly large density range from 0.4 to 1.6, i.e., 0.6 for both electron and hole doping. The value\py{s} of $U_{dd}=6.0$ and $\epsilon_p=3.0$ are set to be relevant for cuprates and Fig.~\ref{LDOS}(d) and (e) further explore the influence of $U_{pp}$. 

At half-filling, the system is \py{insufficient} to open a complete energy gap near \py{the} Fermi level due to relatively low $U_{dd}$ and $\epsilon_p$ as well as our moderate simulated temperature scale. The first noticeable feature is the broad blue O band near -4 eV. To the left and right of the Fermi level lie the ZRS and UHB, respectively, \py{in spite of the presence of a two-peak structure in the UHB}. Limited by our relatively high temperature scale, \py{the so-called Zhang-Rice triplet and the lower Hubbard band~\cite{kung_characterizing_2016, fulde_electron_1995} are} thermally broadened so that merge into the background.

When holes are doped into the system, the broad O band and UHB both shift to the right linearly with hole doping, whereas the ZRS remains close to the Fermi level. On the other hand, electron doping shifts the O band and UHB to the left and rapidly suppresses the `ZRS' to being barely visible. Owing to the chemical potential shift, both peak features near the Fermi level are now interpreted as part of the UHB. The low-energy spectrum develops a splitting feature at a hole doping of $\sim 0.2$, which evolves into more and more complex structures e.g. at $\sim 0.6$ doping. 
Notice that not only the single ZRS peak at small dopings splits into many irregular peaks; but also part of the O spectra shift toward a higher energy, which hints as the breakdown of ZRS at \py{heavily} hole doped systems~\cite{kim_optical_2021}. At the electron doping side of $\sim 0.6$ doping, the spectra \py{show} no anomalous feature which retains high Cu-O hybridization.
As one of our major findings, the possible ZRS breakdown is consistent with
recent experiment~\cite{yin_private}, which showed signatures of an additional O K-edge excitation above the Fermi level in extremely overdoped $\mathrm{La}_{2-x}\mathrm{Sr}_x\mathrm{CuO}_4$ (up to $x = 0.6$). 

Next we display the role of $\epsilon_p$ and $U_{dd}$ in modifying the LDOS for a selected hole doping level of 0.2 in Fig.~\ref{LDOS}(b) and (c), respectively. In Fig.~\ref{LDOS}(b), $U_{dd}$ is fixed at 6.0 while $\epsilon_p$ varies from 2.0 to 6.0; in Fig.~\ref{LDOS}(c), $\epsilon_p$ is fixed at 4.0 while $U_{dd}$ varies from 4.0 to 8.0. Both increasing $\epsilon_p$ and $U_{dd}$ effectively enlarge the distance between the UHB and the broad O band, with the effect of $\epsilon_p$ being significantly more pronounced which reflects the CTI nature of our undoped model. Interestingly, either increasing $\epsilon_p$ or decreasing $U_{dd}$ can enhance the splitting near the Fermi level. As shown in Fig.~\ref{nCu_nO}(a), large $\epsilon_p$ induces more deviated Cu-O hole distribution, \py{which} restricts the stability of ZRS and leads to the peak splitting in Fig.~\ref{LDOS}(b), although a strong $\epsilon_p$ enlarges the effective repulsion on Cu to promote the localized behavior. The same reasoning applies for small $U_{dd}$ by combining Fig.~\ref{nCu_nO}(b) and Fig.~\ref{LDOS}(c). 

In contrast to 0.6 hole doping \py{level}, the prominent Cu-O hybridization is preserved at 0.2 doping in Fig.~\ref{LDOS}(b) and (c), which is evidenced by the coincidence of the spectral peaks of Cu and O. This may point to a distinct origin of the spectral splitting from the ``ZRS breakdown'' in the heavily overdoped regime shown in Fig. ~\ref{LDOS}(a). Extensive earlier theoretical investigations have established the appearance of a new quasiparticle peak (QP) in the vicinity of the Fermi level induced by hole doping, which is commonly interpreted as a dynamical spectral weight transfer~\cite{sordi2025ambipolar, QP_2010, Wang2010prb, chen_doping_2013}. Specifically, in Fig~\ref{LDOS}(a), our LDOS of the density in the range of $0.6\sim 1.2$ successfully reproduce\py{s} the low\py{-}energy feature illustrated by Moritz \textit{et al.}~\cite{moritz2009effect} in the single-band Hubbard model, although the QP in our LDOS is less coherent due to the high \py{simulation} temperature. This may reflect the validity of the low-energy physics of the single-band model at low doping. 

The splitting feature motivates us to further check the impact of the onsite repulsion $U_{pp}$ on O sites. Due to the limitation of the sign problem, we perform this at $U_{dd}=6.0$ and $\beta = 10.0$. When $U_{pp}$ is taken into account, as illustrated in Fig.~\ref{LDOS}(d) and (e), it significantly pushes the onset doping of low\py{-}energy splitting to a higher level. In Fig.~\ref{LDOS}(e), the low\py{-}energy peak of Cu remains intact at $\langle n_{\mathrm{tot} \rangle} = 1.43$ when $U_{pp} = 2.0$, though there is more pronounced O spectral weight transfer to $\sim1.0$ eV as the doping level \py{increases}. In other words, the ZRS is more robust with hole doping at finite $U_{pp}$.

Moreover, increasing $U_{pp}$ appears to drive the UHB closer to the Fermi level by comparing Fig.~\ref{LDOS}(a) with (e). Regardless of the value of $U_{pp}$, within our parameter range, the UHB in the half-filled LDOS consistently exhibits a double-peak structure. Such behavior may indicate either additional excitations or a nontrivial redistribution of spectral weight. 	On the electron-doped side, there is no clear indication that the spectrum is significantly affected by $U_{pp}$. Since the remaining holes primarily lie on the Cu orbitals in this situation, double occupancy on the O orbitals, which inherently have a lower density due to higher onsite energy, is unlikely to occur. The two low-energy peaks, linked to the UHB, show contrasting behavior with increasing electron doping: the right peak diminishes and the left peak becomes more prominent.

To identify more details \py{of} the spectral features near the UHB at half-filling, Fig.~\ref{LDOS}(f) illustrates the temperature $\beta=1/T$ evolution corresponding to the situation in Fig.~\ref{LDOS}(a). As the temperature cools down (increasing $\beta$), the insulating gap gradually becomes clearer. The strong Cu-O hybridization reflected by the coincidence of their spectral peaks is maintained at each $\beta$ value. Notably, the anomalous features above $\omega=0$ of Cu spectra emerge from $\beta = 8.0$ and persists up to $\beta = 15.0$. Conversely, the O spectra never show any anomalous behavior. 

\subsection{$\mathbf{k}$-resolved spectra: pseudogap feature}

\begin{figure*}[t!]
\psfig{figure=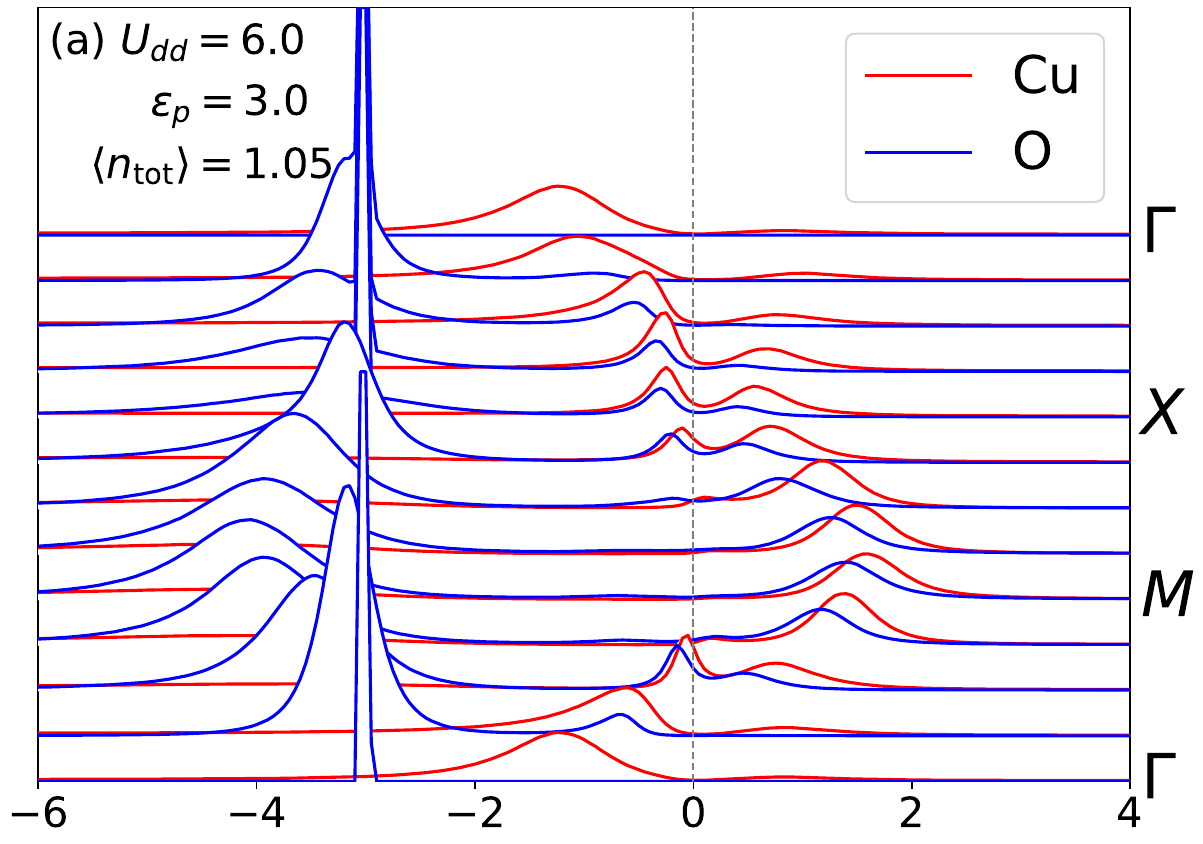,height=4.7cm,width=.45\textwidth,clip}
\psfig{figure=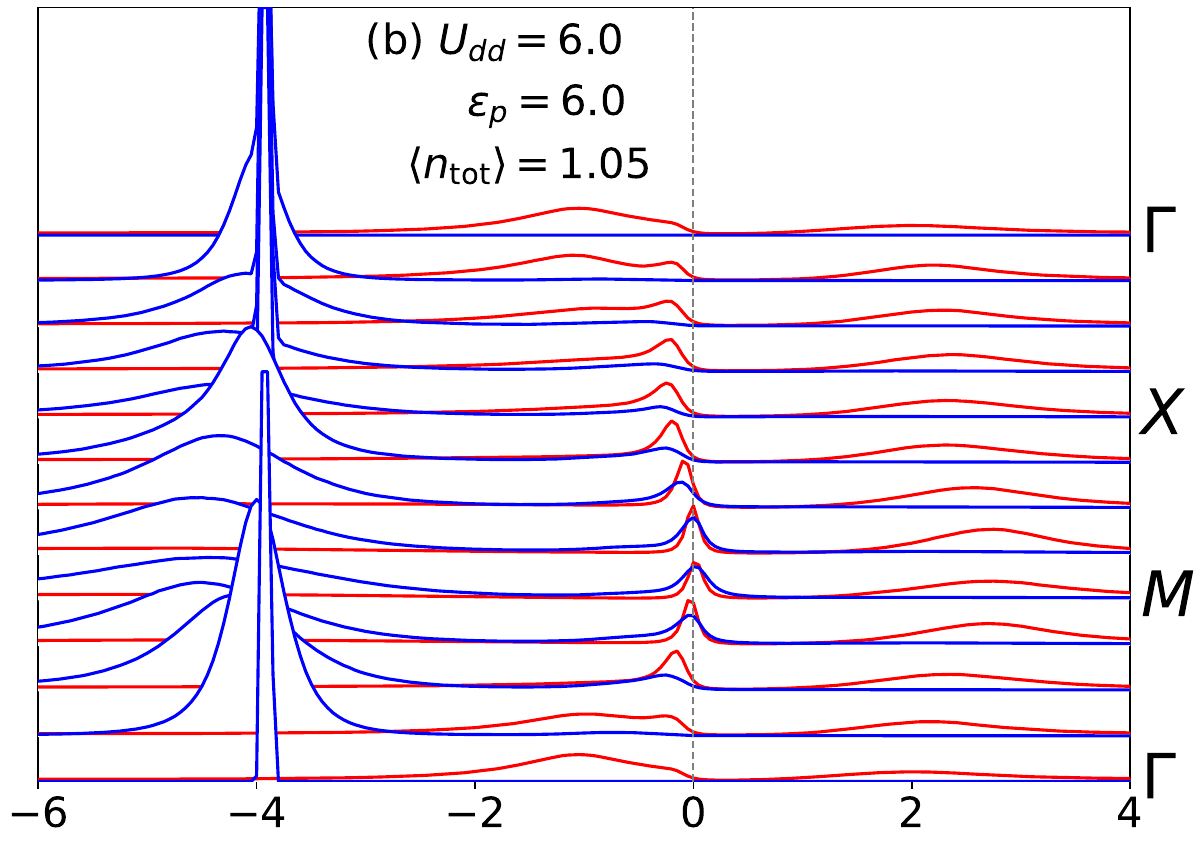,height=4.7cm,width=.45\textwidth,clip}
\psfig{figure=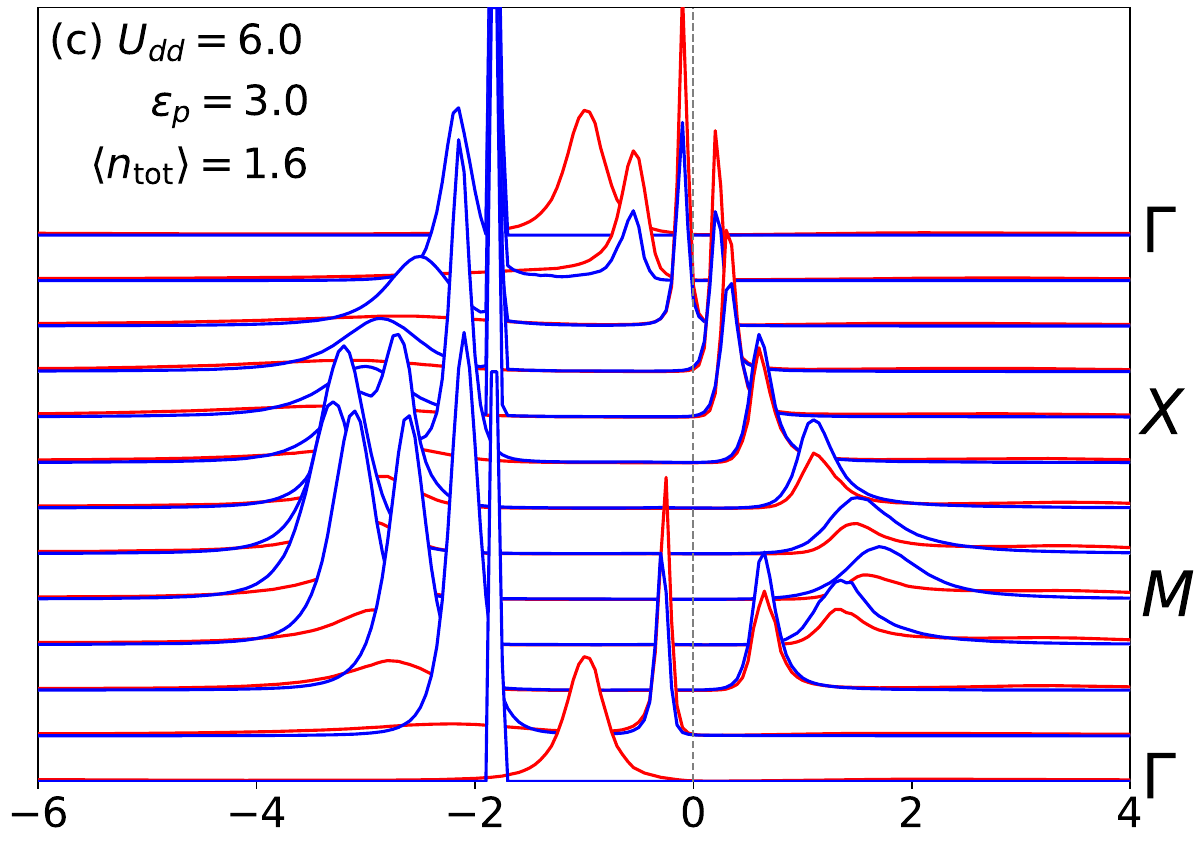,height=4.7cm,width=.45\textwidth,clip}
\psfig{figure=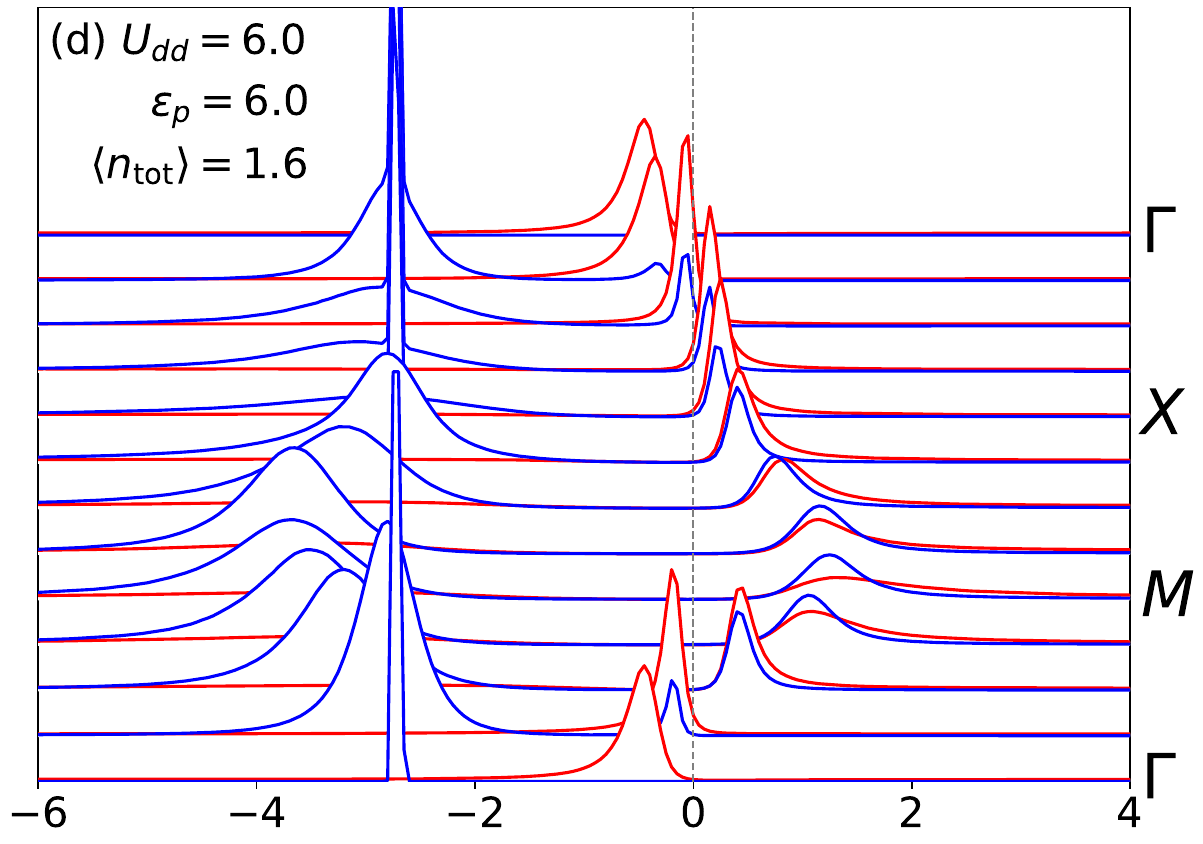,height=4.7cm,width=.45\textwidth,clip}

\caption{The orbital-resolved spectral function $A_{\alpha}(\mathbf{k},\omega)$ along the high-symmetry path $\Gamma$-$M$-$X$-$\Gamma$ in the Brillouin zone. Following the discussion of LDOS, $U_{dd}$ is kept constant at 6.0 as well. The spectral weight of O at $\Gamma$ point is truncated for better clarity. Owing to the momentum-space anisotropy of the spectral function, the O spectrum is obtained by summing the contributions along the $x$- and $y$-directions.}

\label{Akw}
\end{figure*}

The momentum\py{-}dependent single-particle spectral function $A_\alpha(\mathbf{k}, \omega)$ can manifest more information than the local DOS. 
Prompted by the anomalous LDOS near the Fermi level at high doping levels, we further examine the structure of $A_{\alpha}(\mathbf{k}, \omega)$ in Fig.~\ref{Akw}. Here we choose two representative hole doping levels, 0.05 and 0.6, and two typical charge transfer energy scales, $\epsilon_p=3.0 , 6.0$, for our analysis. The first notable feature is the more pronounced spectral weight near the Fermi level along the nodal (N) than the anti-nodal (AN) direction in Fig.~\ref{Akw}(a), which is widely considered to be characteristic of the pseudogap in the underdoped regime of cuprates~\cite{wu_pseudogap_2018, tajima_correlation_2024, vsimkovic2024origin}. 
With the constraint of QMC sign problem, we are unable to reveal a more pronounced pseudogap feature by reducing the temperature so that a decisive conclusion cannot be drawn at present. 
Nevertheless, for $\epsilon_p=6.0$ in Fig.~\ref{Akw}(b), the strong momentum differentiation largely weakens. Specifically, the low-energy peaks are predominantly located around $(\pi, \pi)$ and the spectra do not show any pseudogap feature. 
This difference implies the weakening or absence of the pseudogap in infinite-layer nickelates due to its large charge transfer energy, which also results in lower superconducting $T_c$ than cuprates and the absence of long-range antiferromagnetic order~\cite{mi20prl}.
Additionally, the peak broadening is more evident than the $\epsilon_p=3.0$ case, indicating a larger scattering arising from stronger correlation~\cite{mao_non-fermi-liquid_2024}.

\py{In} the heavily overdoped regime, as illustrated in Fig.~\ref{Akw}(c) and (d), the low-energy peaks become much more coherent and the UHB is hardly visible. 
Compared to the underdoped regime shown in Fig.~\ref{Akw} \py{(a) and (b)}, the difference between panels (c) and (d) is generally less obvious, reflecting the minor role of large $\epsilon_p$ in the heavily overdoped regime. 
\py{Instead, one common feature for $\epsilon_p=3.0$ and $6.0$ is that both cases exhibit stronger zero-energy excitations along the AN direction than along the N direction.} This phenomenon has been widely reported across various cuprates accompanied by the Lifshitz transition of the Fermi surface~\cite{zhong2022differentiated, zhong2018continuous, horio2018three, yong2009dominance}, though the doping level here is much higher. Consistent with the previous research~\cite{kung_characterizing_2016}, the strong Cu-O hybridization feature near the Fermi level persists up to a doping level of 0.6 when $\epsilon_p = 3.0$. However, a notable suppression of hybridization can be identified in Fig.~\ref{Akw}(d). 

Notably, there is no multi-peak feature near the Fermi level at any $\mathbf{k}$-point, suggesting that the multi-peak structure in LDOS of Fig.~\ref{LDOS} originates from momentum integration over distinct regions of the Brillouin zone. 
The additional Cu peaks near the Fermi level in Fig.~\ref{LDOS}(a) at 0.6 hole doping primarily originate around $\Gamma$. On the other hand, the extra O peak above the Fermi level emerges near \py{$\mathrm{M}$}, hence leading to a complicated peak structure near the Fermi level. 

\begin{figure}[h!]
\psfig{figure=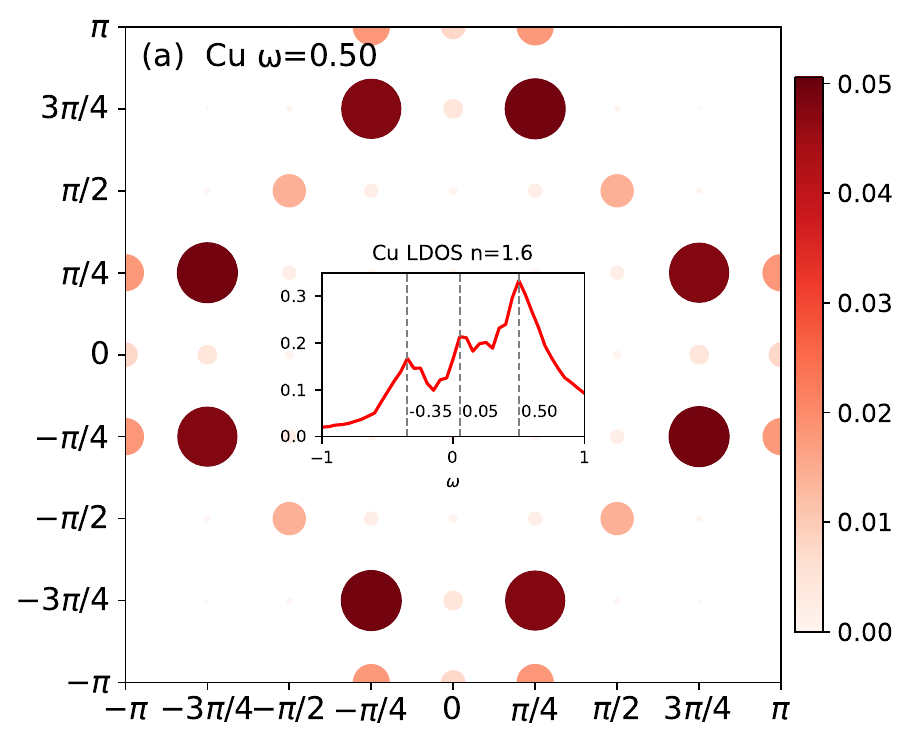,width=.45\textwidth, clip} 
\psfig{figure=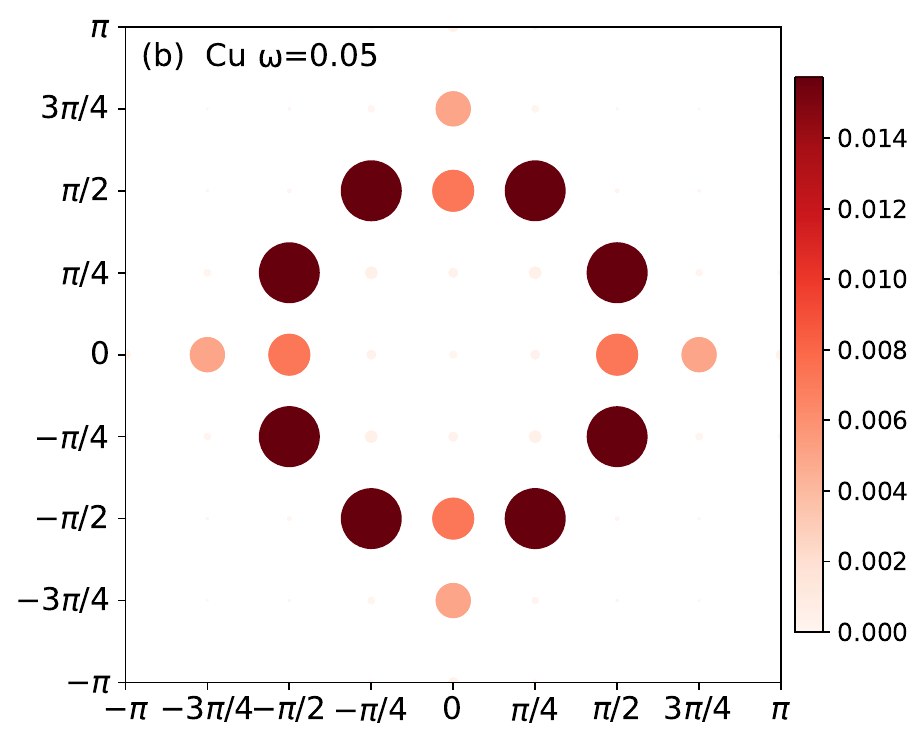,width=.45\textwidth, clip} 
\psfig{figure=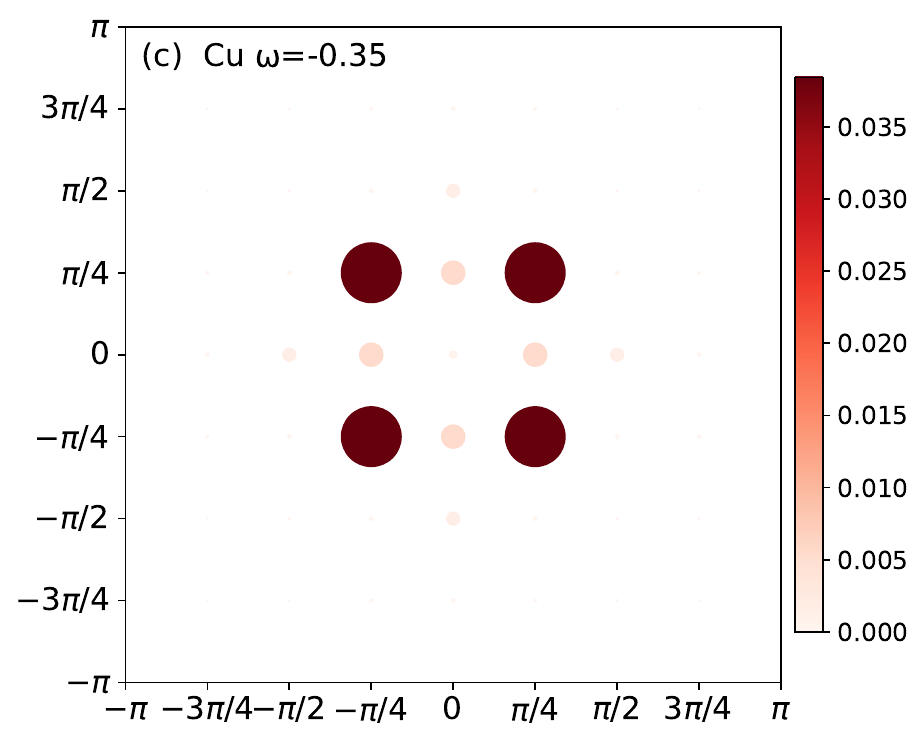,width=.45\textwidth, clip} 
\caption{\py{The contributions of different $\mathbf{k}$ points corresponding to the three spectral peaks (shown in the inset) around the Fermi level of LDOS at 0.6 hole doping in Fig.~\ref{LDOS}(a).}}
\label{k_peaks}
\end{figure}

\py{This phenomenon is reminiscent of the non-interacting dispersion of the Emery model~\cite{tseng2025single}, in which there exists a largely $d$-derived band near the Fermi level when the system is hole-doped. Hence, we extract the low-energy peak positions of Cu in Fig.~\ref{LDOS}(a) for a hole doping of 0.6. Fig.~\ref{k_peaks} shows the contributions of different $\mathbf{k}$ points to these three extracted peaks labeled in the inset of panel (a), with each peak integrated over a window of $\pm 0.05$. These iso-energy surfaces clearly show that the dispersion evolves from electron-like to hole-like as the energy increases from low to high, which corresponds very well to the $d$-feature band structure in the non-interacting limit. At least from the perspective of our Emery model, the intrinsic non-interacting dispersion naturally gives rise to the multi-peak character of the low-energy physics at high doping concentrations.}

\subsection{Magnetic properties}

\begin{figure}
\psfig{figure=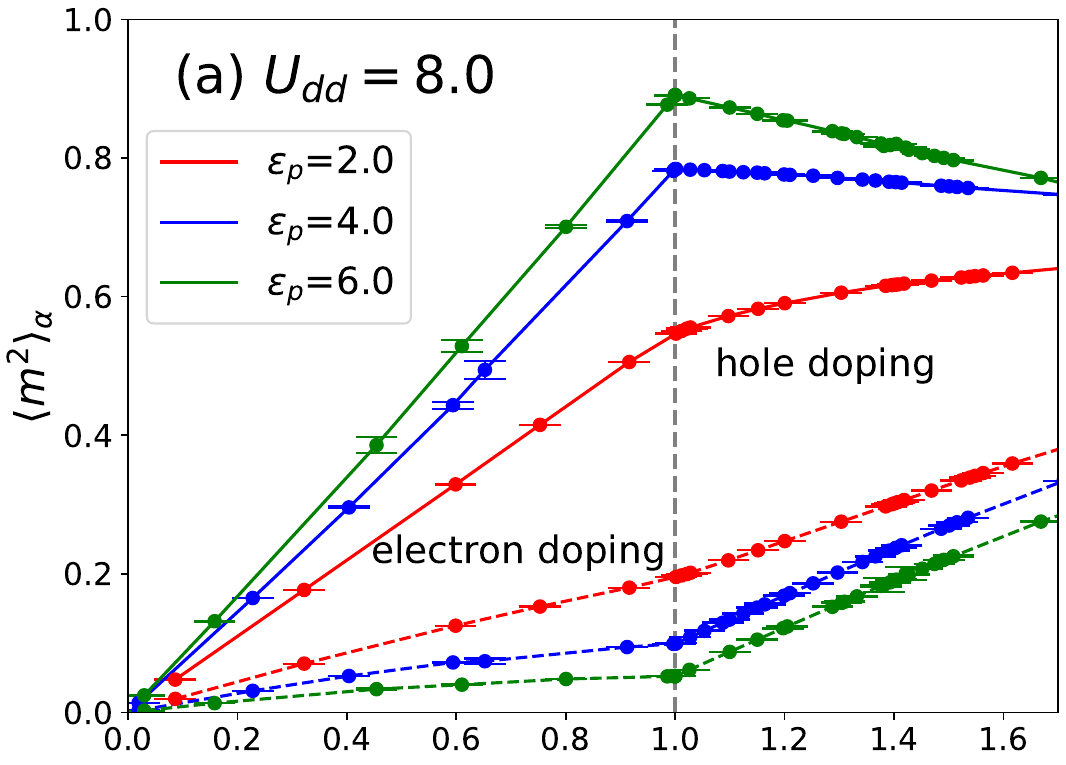,height=5.2cm,width=.45\textwidth, clip} 
\psfig{figure=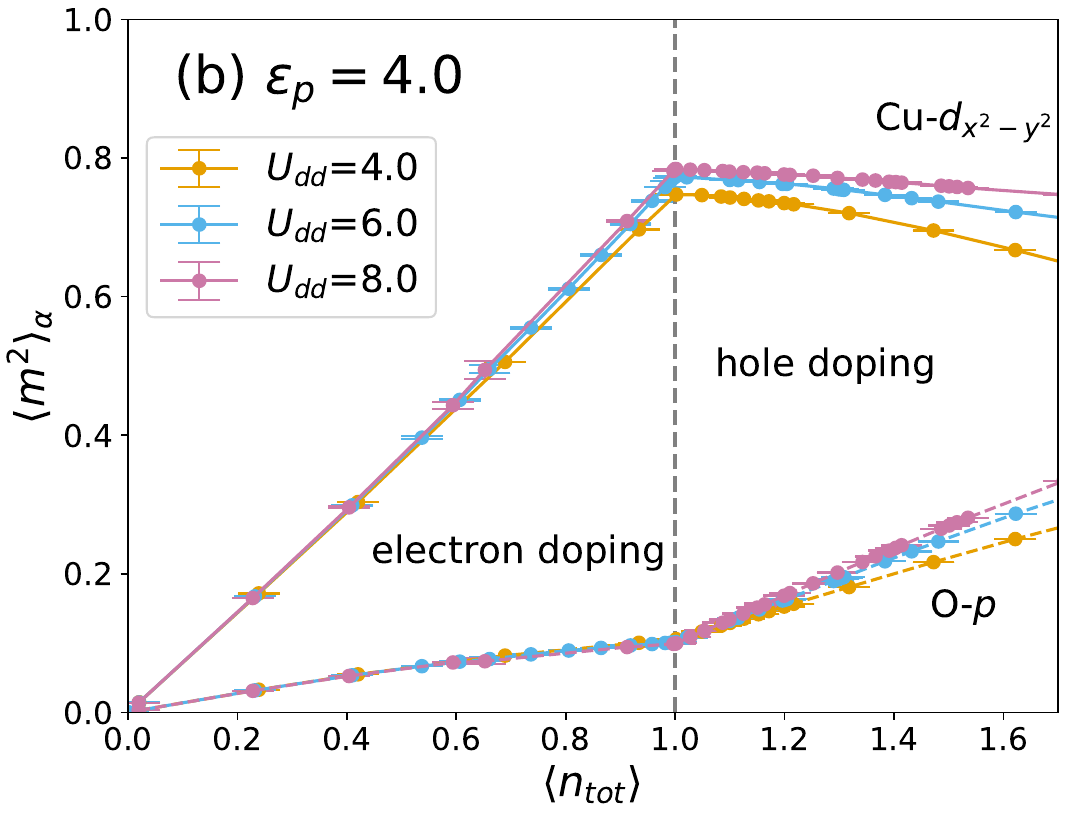,height=5.2cm,width=.45\textwidth, clip} 
\caption{Orbital-resolved local moment $\langle m^2 \rangle_\alpha$ versus $\langle n_{\mathrm{tot}} \rangle$ with $U_{dd}$ or $\epsilon_p$ varies. Here $\alpha=$Cu/O is distinguished by solid/dashed line.}
\label{m2}
\end{figure}

From now on, we concentrate on two-particle quantities such as the spin-spin correlation and the spin structure factor, as both experimental and theoretical studies~\cite{Lichtenstein_AFM_2000, hart2015observation} have revealed N\'eel antiferromagnetic ordering near zero doping. Therefore, a careful examination of how $U_{dd}$ and $\epsilon_p$ influence these quantities is warranted. We first examine the orbital-resolved local moment, $\langle m^2 \rangle_\alpha = \langle (n^\alpha_\uparrow-n^\alpha_\downarrow)^2 \rangle$, which quantifies the localized behavior of spins as a precondition for the emergence of magnetic order. Since $\epsilon_p$ adjusts the distribution of the doped electron/hole directly, one can readily anticipate its strong influence on the local moment.

On the one hand, as illustrated in Fig.~\ref{m2}(a), increasing $\epsilon_p$ \py{clearly} enhances the local moment on Cu orbital \py{, which enlarges the effective interaction on Cu-$d$ orbital}. The local moment at both electron and hole doping sides show nearly linear dependence with strong slope asymmetry. At the hole doping side, the decreasing $\langle m^2 \rangle_{\mathrm{Cu}}$ with doping \py{for $\epsilon_p > 2.0$} suggests that the magnetic correlations would also be weakened. Although $\langle m^2 \rangle_{\mathrm{O}}$ obviously increases at the hole doping side, its amplitude is much smaller compared to Cu and contributes less to the magnetic response in the system, as clearly evidenced in \py{a previous} DQMC study~\cite{kung_characterizing_2016}. On the other hand, Fig.~\ref{m2}(b) indicates \py{a} less significant impact of $U_{dd}$, especially at the electron doping side. The onsite repulsion on the Cu orbital effectively limits the double occupancy and thereby enhances $\langle m^2 \rangle_{\mathrm{Cu}}$ at the hole doping side.

\begin{figure}
\psfig{figure=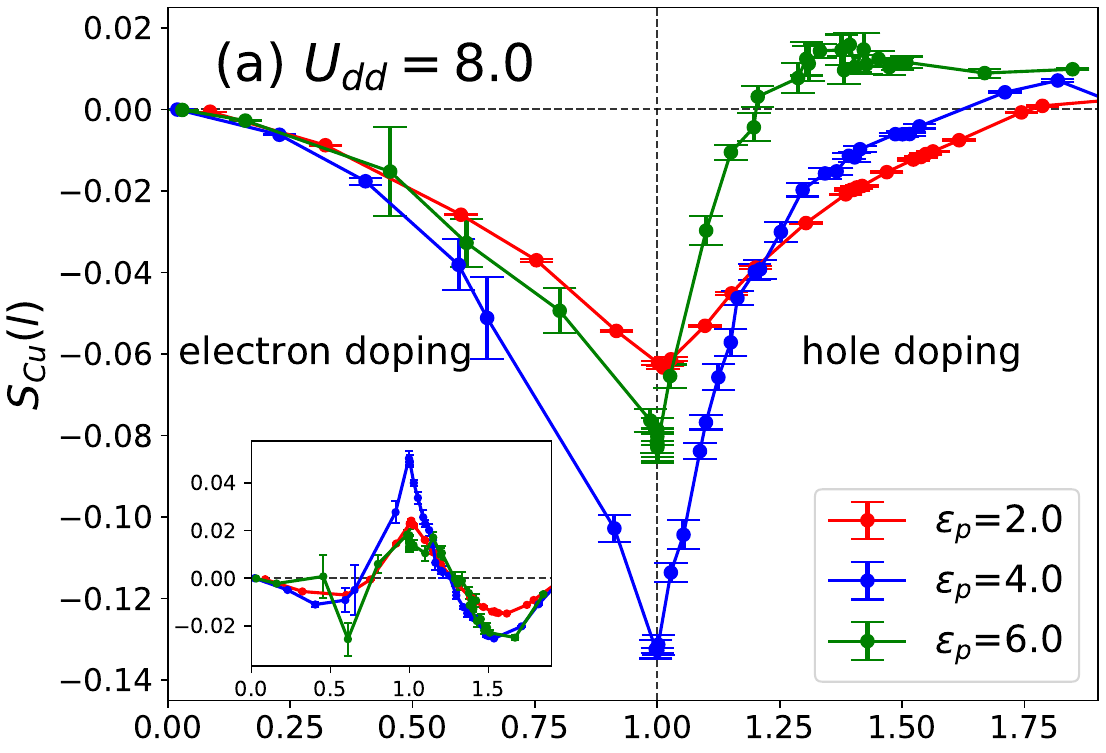,height=5.2cm,width=.45\textwidth, clip} 
\psfig{figure=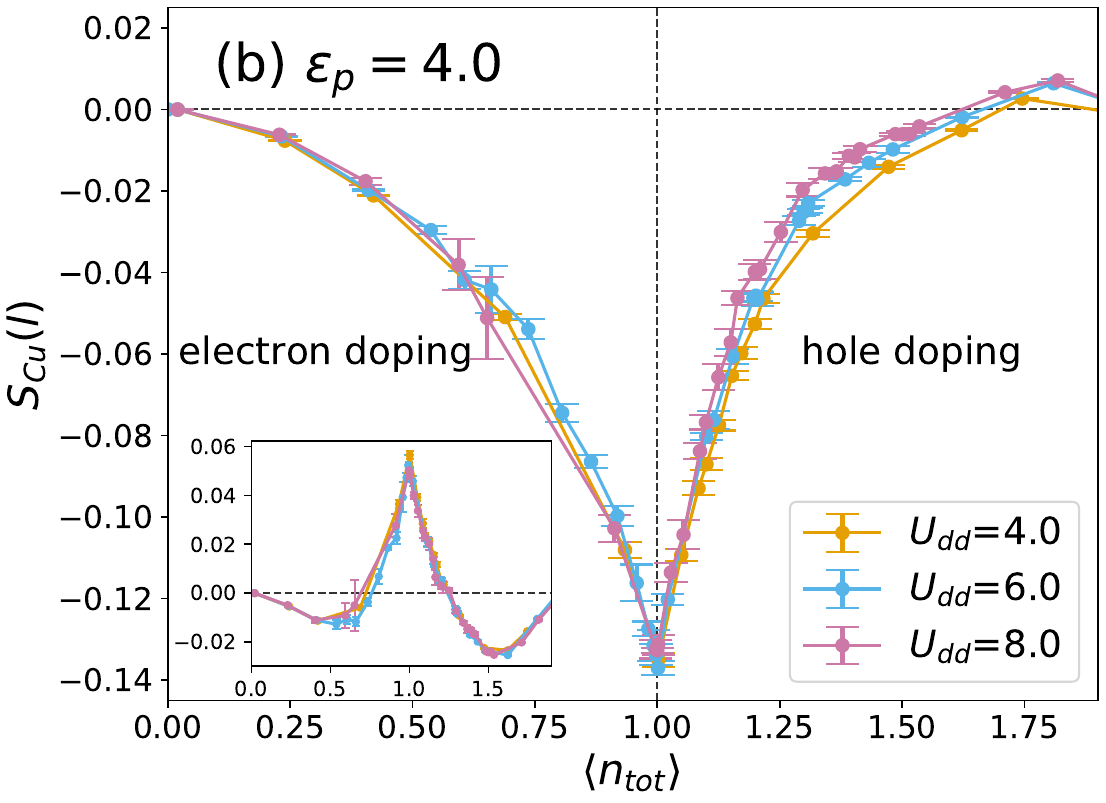,height=5.2cm,width=.45\textwidth, clip} 
\caption{Nearest neighbor spin-spin correlation function $S_{\mathrm{Cu}}(1, 0)$ versus $\langle n_{\mathrm{tot}} \rangle$ with $U_{dd}$ or $\epsilon_p$ varying. The inset shows the next-nearest neighbor $S_{\mathrm{Cu}}(1, 1)$. \py{The horizontal dashed line is an indicator of sign change. The vertical dashed line separates hole or electron doping.}}
\label{spinr}
\end{figure}

\begin{figure}
\psfig{figure=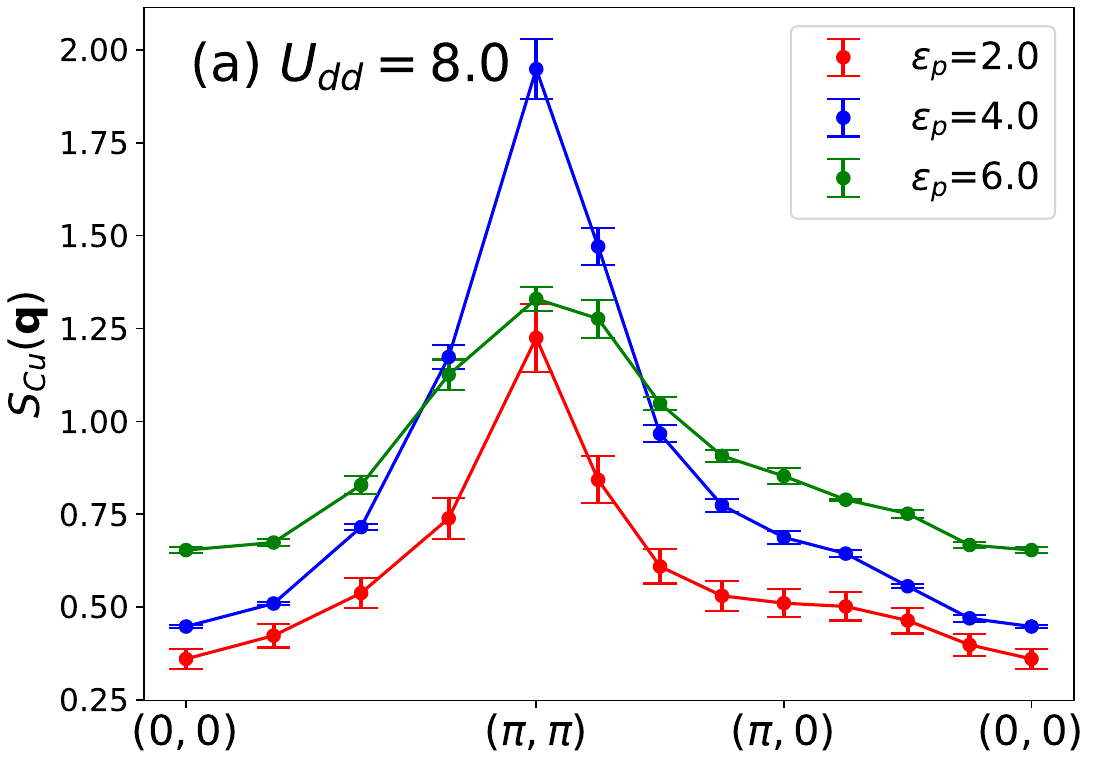,height=5.2cm,width=.45\textwidth, clip} 
\psfig{figure=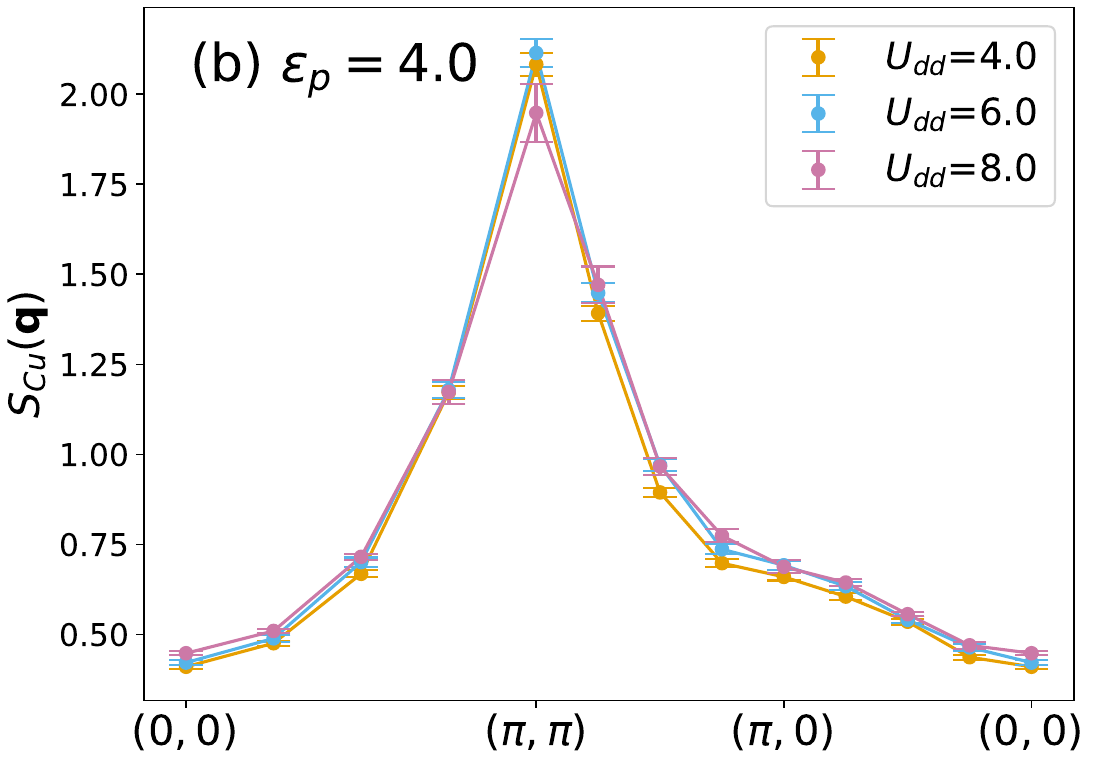,height=5.2cm,width=.45\textwidth, clip} 
\psfig{figure=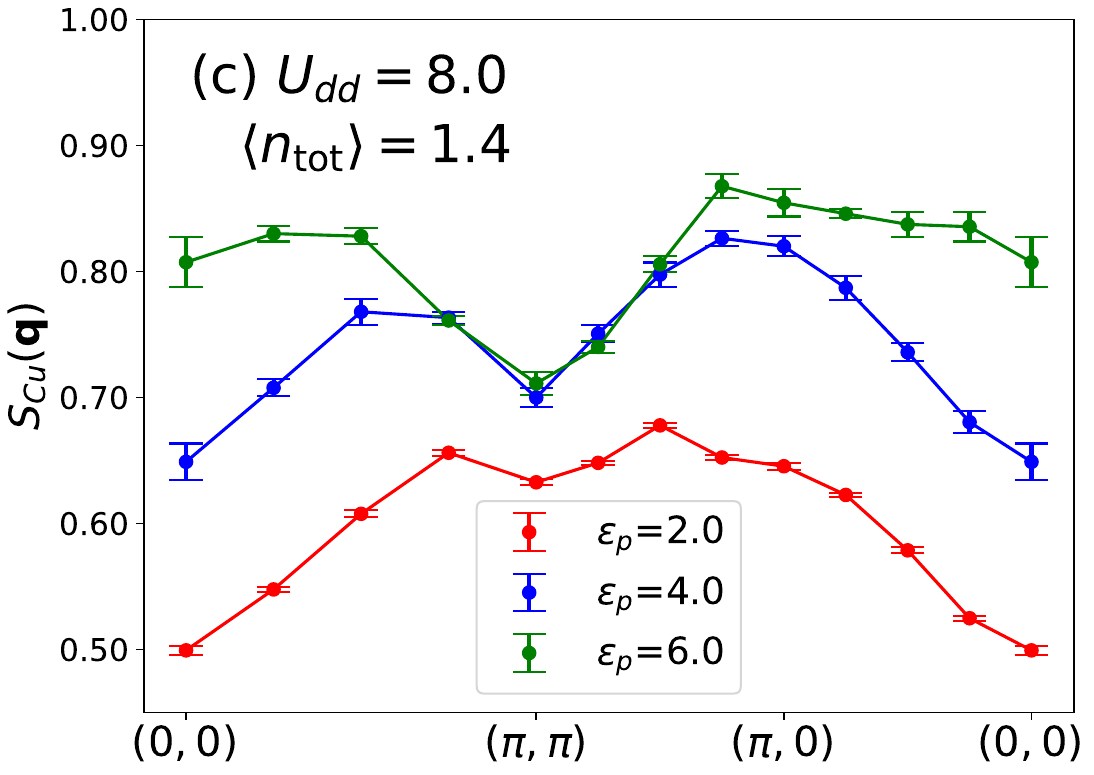,height=5.2cm,width=.45\textwidth, clip} 
\caption{Cu-Cu spin structure factor $S_{\mathrm{Cu}}(\mathbf{q})$ with $\mathbf{q}$ along the high-symmetry path $\Gamma$-$M$-$X$-$\Gamma$ in the first Brillouin zone for various values of $U_{dd}$ or $\epsilon_p$: panels (a) and (b) correspond to half-filling, while panel (c) corresponds to 0.4 hole doping. }
\label{spink}
\end{figure}

In our SU(2)-symmetric system under investigation, the complete spin rotational symmetry is maintained. Nonetheless, numerically z-component quantities show less uncertainty so that we adopt the z-component of spin-spin correlation function 
\begin{align}
     S_\alpha(\mathbf{l}) = \frac{1}{N} \sum_i \left\langle \left( n^\alpha_{i\uparrow} - n^\alpha_{i\downarrow} \right) \left( n^\alpha_{i+\mathbf{l},\uparrow} - n^\alpha_{i+\mathbf{l},\downarrow} \right) \right\rangle
\end{align}
and the static spin structure factor
\begin{align}
S_{\alpha}(\mathbf{q}) = \sum_\mathbf{l} e^{i\mathbf{q} \cdot \mathbf{l}}S_\alpha(\mathbf{l})  
\end{align}
to study how the magnetism evolves with respect to various parameters. The dominant spin-spin correlation of our model is in Cu-Cu channel, especially the short range part $S_{\mathrm{Cu}}(1, 0)$ and $S_{\mathrm{Cu}}(1, 1)$, where $(1,0)$ and $(1,1)$ denote the spatial separation.

Fig.~\ref{spinr}(a) shows $S_{\mathrm{Cu}}(1, 0)$ and $S_{\mathrm{Cu}}(1, 1)$ versus total hole filling. The increasing slope difference between each side of half-filling with increasing $\epsilon_p$ indicates the intrinsic electron–hole asymmetry caused by charge transfer energy of our model. Another notable feature is that the larger $\epsilon_p$ causes a smaller critical doping, at which $S_{\mathrm{Cu}}(1, 0)$ changes its sign to positive, suggesting the emerged short-range ferromagnetic correlation~\cite{jia2014persistent, lau2011high, lau2011computational, ebrahimnejad2014dynamics}.
The inset shows the next-nearest neighbor $S_{\mathrm{Cu}}(1, 1)$. 

Interestingly, in a wide doping range around half-filling, both $S_{\mathrm{Cu}}(1, 0)$ and $S_{\mathrm{Cu}}(1, 1)$ reach their maximum value when $\epsilon_p$ equals 4.0. This feature implies the existence of an optimal charge transfer energy scale, which is reminiscent of an earlier study identifying an optimal $\epsilon_p$ for the maximal superconducting $\mathrm{T_c}$~\cite{mai_pairing_2021}. 

In addition, Fig.~\ref{spinr}(b) indicates that varying $U_{dd}$ does not obviously affect the amplitude of the short range magnetic correlation at all doping levels. Both the peak of $(1, 1)$ and the valley of $(1, 0)$ show strong antiferromagnetic order at half-filling. 

The sign change feature reflecting the transition from antiferromagnetic to ferromagnetic neighboring correlations motivates us to further investigate the parameter dependence of the Cu-orbital's spin structure factor $S_{\mathrm{Cu}}(\mathbf{q})$. 
In Fig.~\ref{spink}(a), the antiferromagnetic ordering vector $\mathbf{q}=(\pi, \pi)$ dominates at half-filling as expected. Similar to the spin-spin correlation function, $S_{\mathrm{Cu}}(\mathbf{q})$ exhibits a maximal antiferromagnetic peak at an intermediate value of $\epsilon_p$. The existence of an optimal $\epsilon_p$ is supported by numerous computational methods~\cite{cui2020ground, mai_pairing_2021, zegrodnik2019superconductivity}. Away from the antiferromagnetic wave-vector, the existence of optimal $\epsilon_p$ gradually disappears and the intensity exhibits the monotonic rise as $\epsilon_p$ increases, which indicates enhanced spin fluctuations and that the Cu orbitals are more localized.  
Again Fig.~\ref{spink}(b) reveals the minor modification from $U_{dd}$. In fact, this insensitivity to $U_{dd}$ is naturally expected since the local moment is robust against varying $U_{dd}$ in Fig.~\ref{m2}(b) as well as the robustness of neighboring spin correlations in Fig.~\ref{spinr}(b).

When the hole doping level reaches 0.4 and the optimal $\epsilon_p$ behavior has disappeared in Fig.~\ref{spinr}(a), Fig.~\ref{spink}(c) presents the corresponding spin structure factor. Because of the increasing local moment in Fig.~\ref{m2}(a), the overall intensity is shifted up by enlarging $\epsilon_p$.
Meanwhile, the AFM\py{-}dominated peak gradually diminishes and the $\mathbf{q}$ distribution \py{shifts toward the $\Gamma$ point, suggesting} possible weak incommensurate magnetic ordering. Consistent with the spin correlations, there is an enhancement of short-range ferromagnetic (FM) correlations with increasing $\epsilon_p$. These phenomena at high hole doping may indicate an important role of paramagnon~\cite{wang2022paramagnons, le2011intense, para2007prl} in describing heavily doped systems.

Finally, we provide a brief interpretation for this optimal $\epsilon_p$ behavior. Consistent with Cui \textit{et al.}~\cite{cui2020ground}, the local magnetism is naturally strengthened at high $\epsilon_p$ because of effectively enlarged $U_{dd}$. The low AFM correlation at small $\epsilon_p$ is clear because of the insufficient local moment on Cu orbitals. Nonetheless, at large $\epsilon_p$ values, the superexchange given in four\py{th}-order perturbation as~\cite{YU1998137, zhang_effective_1988, blankenbecler_monte_1981, zaanen1988, eskes1993, mi20prl}
\begin{align}
    J = \frac{4t_{pd}^4}{\Delta^2} \left( \frac{1}{\Delta} + \frac{1}{U_{dd}} \right)
\end{align}
decreases monotonically with increasing $\epsilon_p$. At $\epsilon_p=6.0$, the remaining superexchange is only about 20\% of the typical value in cuprates. The much weaker superexchange results in the diminished spin correlations and structure factor in spite of a stronger local magnetic moment. Hence, the competition between the effective magnetic moment and the superexchange leads to the maximum AFM correlation for a moderate $\epsilon_p$. Since the unconventional SC is widely considered to be closely associated with AFM correlations~\cite{mukuda2010superexchange, foley2019coexistence, plakida2006theory}, our results may offer some insights for adjusting the magnetism to optimize the superconducting $T_c$.

\begin{figure}
\psfig{figure=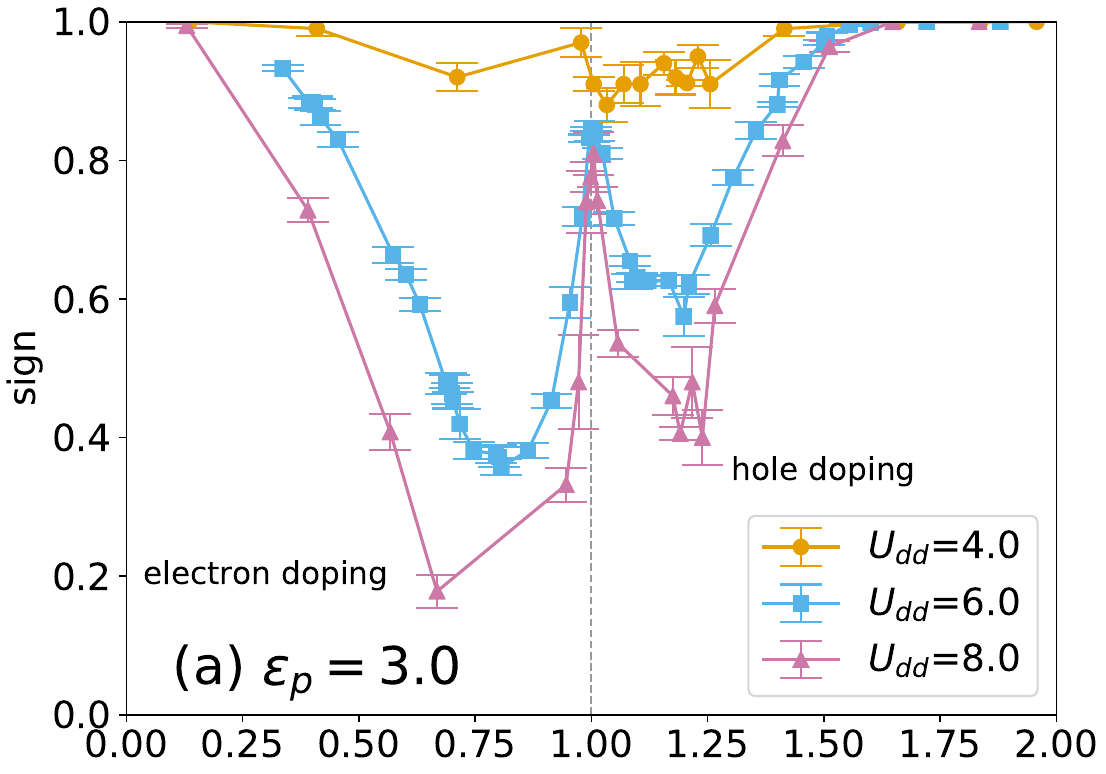,height=5.2cm,width=.45\textwidth, clip} 
\psfig{figure=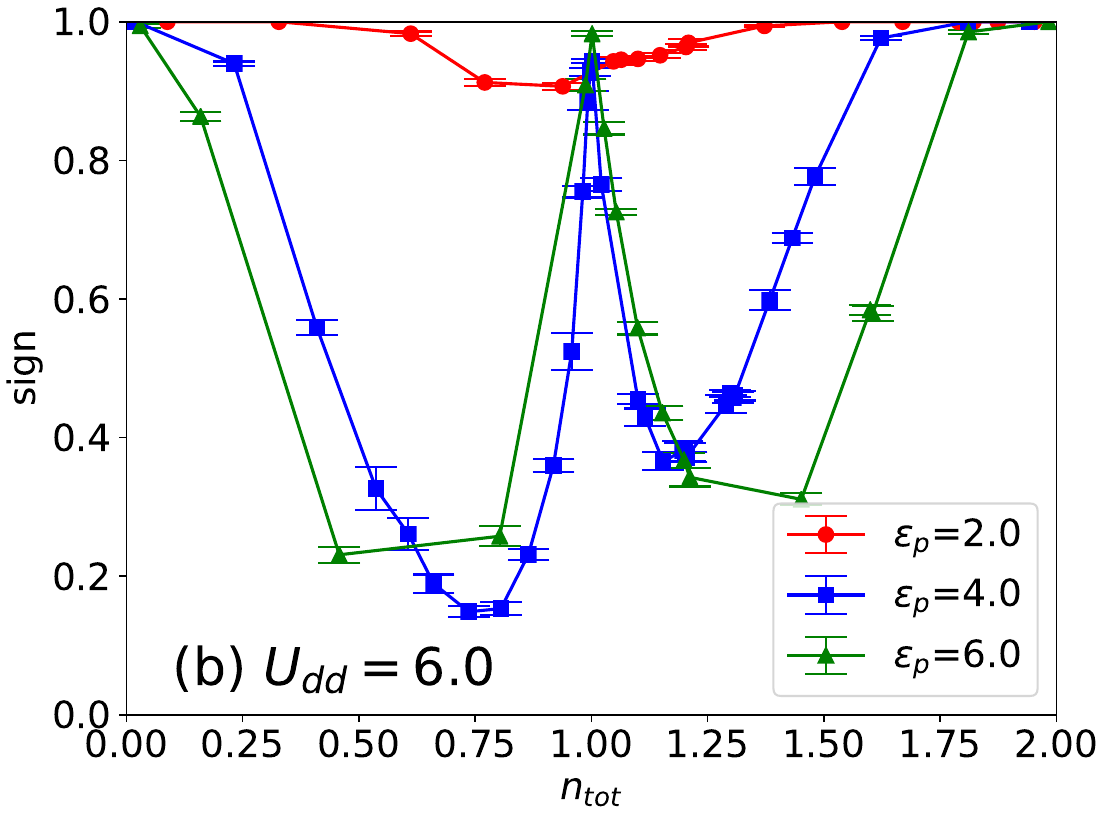,height=5.2cm,width=.45\textwidth, clip} 
\caption{Average fermion sign as a function of total filling $\langle n_{\mathrm{tot}} \rangle$. All the simulations shown are performed on an $8 \times 8$ lattice with $\beta = 10$.}
\label{sign}
\end{figure}

\section{Conclusion}
\label{sec:conclusion}

In summary, by employing large-scale DQMC simulations, we have investigated the influence of $U_{dd}$, $\epsilon_p$, $U_{pp}$, as well as the electron/hole doping in the three-orbital Emery model on physical quantities such as orbital occupancy, local and k-resolved spectral functions, as well as spin correlation functions. 

We concentrate on the difference arising from the large $\epsilon_p$ that is believed to be relevant to infinite-layer nickelate superconductors~\cite{mao_non-fermi-liquid_2024, mi20prl}. The pseudogap features at small charge transfer energy scale (relevant to cuprates) are shown to diminish at larger $\epsilon_p$, which implies the weakening or absence of the pseudogap in the infinite-layer nickelates.
In addition, the spectra of low doping levels are basically consistent with the characteristic of dynamical spectral weight transfer. However, signatures of ZRS breakdown have been identified via the spectra in the heavily overdoped regime. This undoubtedly challenges the applicability of the single-band Hubbard model. The anomalies in the spectral function motivate our further investigation of magnetism. Around the antiferromagnetic wave vector $(\pi, \pi)$, an optimal $\epsilon_p\sim 4.0$ which gives rise to the largest spin correlation and structure factor near the half-filling is detected, which is closely related to the superexchange mechanism in cuprates. At high doping levels, a higher $\epsilon_p$ results in a stronger short-range FM fluctuation. 

All these findings above highlight the pivotal role of the charge transfer energy in shaping both the spectral and magnetic responses, and shed light on the applicability of the three-orbital Emery model as a common framework for capturing the intertwined phenomena in cuprates and infinite-layer nickelates~\cite{mao_non-fermi-liquid_2024}. In the CTI regime, a negligible role of the onsite interactions on investigating the magnetism is revealed, which may significantly mitigate the sign problem in the future.


\section{Appendix: QMC sign problem}

\py{Fig.~\ref{sign} (a) and (b) show the average fermionic sign for most parameter sets used in this study. When $U_{pp}$ is zero, the fermionic sign is partially relieved particularly at half-filling~\cite{kung_characterizing_2016}. The most severe sign problem occurs in the intermediate doping regime for both electrons and holes. For the commonly used parameter set $U_{dd}=6.0$ and $\epsilon_p=3.0$ in the discussion of the spectral function, the average sign drops to around 0.4, which remains reasonably high. Therefore, we normally perform 5000 thermalization sweeps, followed by calculations, divided into 10 bins with 10000 measurements each.}

\section{Acknowledgments}
We would like to thank George A. Sawatzky, Xinmao Yin, Xiongfang Liu, Rui Peng, and Wenxin Ding for illuminating discussions.
We acknowledge the support by National Natural Science Foundation of China (NSFC) Grant No.~12174278.

\bibliography{main}



\end{document}